\documentclass[useAMS,usenatbib,usedcolumn]{mn2e}
\usepackage{multirow}
\usepackage{amsmath}
\usepackage{graphicx,epsfig}
\usepackage[graphicx]{realboxes}
\usepackage{float}
\usepackage[colorlinks=true,citecolor=blue]{hyperref}%

\RequirePackage{rotating}
\title[A Multi-Membership OC Catalogue]{A Multi-Membership Catalogue for 1876 Open Clusters using UCAC4 data.}

\author[Sampedro et al.]
{L. Sampedro$^{1,2}$, W. S. Dias$^3$, E. J. Alfaro$^2$, H. Monteiro$^3$, A. Molino$^{1,2}$
\\$^1$ Instituto de Astronom\'ia, Geof\'isica e Ci\^encias Atmosf\'ericas, Universidade de S\~ao Paulo, Cidade Universit\'aria, S\~ao Paulo SP, Brazil
\\e-mail: lauramariash@gmail.com
\\$^2$Instituto de Astrof\'isica de Andaluc\'ia. IAA-CSIC. Glorieta de la Astronom\'ia S/N. 18008, Granada. Spain
\\$^3$UNIFEI, Instituto de F\'isica e Qu\'imica, Universidade Federal de Itajub\'a, Av. BPS 1303 Pinheirinho, 37500-903 Itajub\'a, MG, Brazil}
\date{Submitted to MNRAS}
\date{Accepted XXX. Received YYY; in original form ZZZ}

\begin{document}
\label{firstpage}

\pagerange{\pageref{firstpage}--\pageref{lastpage}} 
\maketitle{}

\begin{abstract}

The main objective of this work is to determine the cluster members of 1876 open clusters, using  positions and proper motions of the astrometric catalogue UCAC4. For this purpose we apply three different methods, all them based on a Bayesian approach, but with different formulations: a purely parametric method, another completely non-parametric algorithm, and a third, recently developed by Sampedro \& Alfaro, using both formulations at different steps of the whole process. The first and second statistical moments of the members phase-space  subspace, obtained after applying the three methods, are compared for every cluster. Although, on average, the three methods yield similar results, specific differences between them, as well as for some particular clusters, are also present. The comparison with other published catalogues shows good agreement. We have also estimated for the first time the mean proper motion for a sample of 18 clusters. The results are organized in a single catalogue formed by two main files, one with the  most relevant information for each cluster, partially including that in UCAC4, and the other showing the individual membership probabilities for each star in the cluster area. The final catalogue, with an interface design that enables an easy interaction with the user, is available in electronic format at SSG-IAA (http://ssg.iaa.es/en/content/sampedro-cluster-catalog) website.

\end{abstract}


%**********************************************************************
%**********************************************************************
\section{Introduction}
\label{intro}
%**********************************************************************
%**********************************************************************
Open clusters are excellent laboratories for exploring a large number of astrophysical problems: they are crucial to set constraints to the stellar evolutionary  models \citep[i.e.][and references therein]{2006A&A...456..269L, 2017IAUS..316É.1C}, and to study in detail the star-formation process  and the possible physical mechanisms driving and controlling all the different steps that lead from a molecular cloud to a gravitational bound set of stars \citep[i.e.,][]{1994LNP...439...13L, 2000prpl.conf..179E, 2010RSPTA.368..713L}. Since the pioneering work by \citet{Becker64}, open clusters have also been  used as suitable probes for drawing  the Galactic structure in terms of  shape, morphology,  kinematics and chemical distribution \citep[i.e.][]{1980A&A....88..360V, 1988AJ.....95..771J, 1991ApJ...378..106A, 2005ApJ...629..825D, 2008AJ....136..118F, 2009A&A...494...95M, 2011MNRAS.417..698L, 2012Msngr.147...25G, 2013A&A...557A..14O, 2013MNRAS.432.3349C, 2014A&A...561A..93H, 2015MNRAS.449.2336J, 2016arXiv161104398C, 2017arXiv170300762M}. Descending to conceptually smaller scales, open clusters are the observational basis of our knowledge on the Initial Mass Function  \citep{2001MNRAS.322..231K, 2010ARA&A..48..339B}, the existence or not of a primordial mass segregation, \citep{1998MNRAS.295..691B, 2009MNRAS.395.1449A, Parker11, Parker17}, the time scales for the destruction or dilution of these systems \citep{Parker09, 2014prpl.conf..243K, 2016MNRAS.461.2519V}, the concurrence of different star-formation bursts in a single cluster \citep{2017ApJ...835...60W}, etc. In short, open clusters represent unique observational targets for the development of Galactic and Stellar Physics \citep[see][]{2012Msngr.147...25G}.

However, in most cases, making use of these stellar systems requires the preceding step of determining the physical members of the cluster. The classification  between members and non-members of the stars located in the stellar field of the cluster depends on how we have defined what a stellar cluster is and which of its properties we use to carry out the classification, which is at the same time influenced by the observed variables available. Here we are using the UCAC4 data \citep{2013AJ....145...44Z} ---that is, positions and proper motions --- hence we must use those phenomenological aspects of the actual definition of cluster that leave their mark on the phase-space subspace formed by these variables. It is evident that the first clue of the existence of a stellar cluster is revealed by an increase in superficial stellar density in the plane of the sky, and since the 1930s we have also known that the kinematic behavior of a cluster is reflected in a greater density, with regard to the field stars, in the space of the kinematic variables. If, moreover, as in this case, the aim is to carry out an analysis of the whole system of Galactic clusters, we also need all-sky, accurate and homogeneous data.

Several efforts have been made over recent years to generate homogeneous and systematic astrometric catalogues, and to use them to analyse the membership to the system of open clusters of the catalogued stars. Examples of these works are: \citet{2000A&AS..146..251B, 2003ARep...47....6L,2004A&AT...23..103B, 2006A&A...446..949D, 2014A&A...564A..79D, 2005A&A...438.1163K, 2012A&A...543A.156K, 2013A&A...558A..53K}. Most of the techniques used to separate field and cluster populations address the problem from a statistical point of view, by computing the membership probabilities through the estimation of the probability density functions (PDFs) for cluster and stellar-field populations in the subspace formed by the astrometric variables. However, the model underlying the distribution of these variables is not unique, and neither is the algorithm used to determine the PDFs and to estimate the cluster membership probabilities. 

Since the pioneering work by \cite{1958AJ.....63..387V} using only proper motions, many other studies have proposed different approaches to the problem that respond to the observational fact of the distribution pattern diversity of the astrometric variables in the clusters studied  \citep[i.e.,][among others]{1985A&A...150..298C, 1990A&A...235...94C, 1990A&A...237...54Z, 1998A&A...337..125G, 2014A&A...564A..79D, 2016MNRAS.457.3949S, 2016RAA....16..184G}. Cluster membership determination is therefore dependent on both the characteristics of the data and the distinct statistical approaches assumed by the different methods. In this work, we apply three different methods fully described in \citet{1985A&A...150..298C, 1990A&A...235...94C} and \citet{2016MNRAS.457.3949S}, to determine, in a homogeneous way, the members of the open clusters listed in \citet{2002A&A...389..871D}. 

The paper is organized as follows. Section  2 summarizes used in this work. In section 3 we make a brief description of the three methods applied to determine the membership of the stars. Section 4 summarizes the main results and discuss the comparison with previous published catalogues. Finally, in Section 5 we highlight the main conclusions.

%**********************************************************************
%**********************************************************************
\section{Data}
\label{data}
%**********************************************************************
%**********************************************************************

We use  version 3.5 of the \textit{New Optically Visible Open Clusters and Candidates catalogue} \citep[][hereafter DAML02]{2002A&A...389..871D},  to select a sample of 2167 open clusters to be analysed. The stellar positions and the proper motions are taken from the \textit{Fourth United States Naval Observatory CCD Astrograph Catalogue}, UCAC4 \citep{2013AJ....145...44Z}. The catalogue contains data for over 113 million stars (105 million of them with proper-motion data) being complete down to magnitude R = 16. The positional accuracy of the listed objects is about 15 - 100 mas per coordinate, depending on the magnitude. Formal errors in proper motions range from about 1 to 10 mas yr$^{-1}$ depending on the magnitude and the observational history. Systematic errors in the proper motions are estimated to be about 1 - 4 mas yr$^{-1}$.

As proposed by \citet{2010A&A...510A..78S}, we should utilize the cluster angular radius as the best sampling radius to analyse the cluster membership when using the  parametric model of the proper-motion distribution. However, given that we are dealing with three different methods we proceed in a different way. We download the data from UCAC4, using the VizieR service\footnote{http://vizier.u-strasbg.fr/viz-bin/VizieR} \citep{2000A&AS..143...23O} for an initial cluster centre and within an area given by the DAML02 catalogued radius plus extra 15 arcminutes to ensure we take all the stars in the cluster region. Then we again calculate apparent cluster angular radii through visual inspection of the radial density profiles (RDPs), where the apparent cluster angular radius is defined as the distance from the cluster centre where the RDP drops into the field. We should note that for this estimation we only make use of the stars catalogued in UCAC4.

\begin{figure}
\includegraphics[width=8cm]{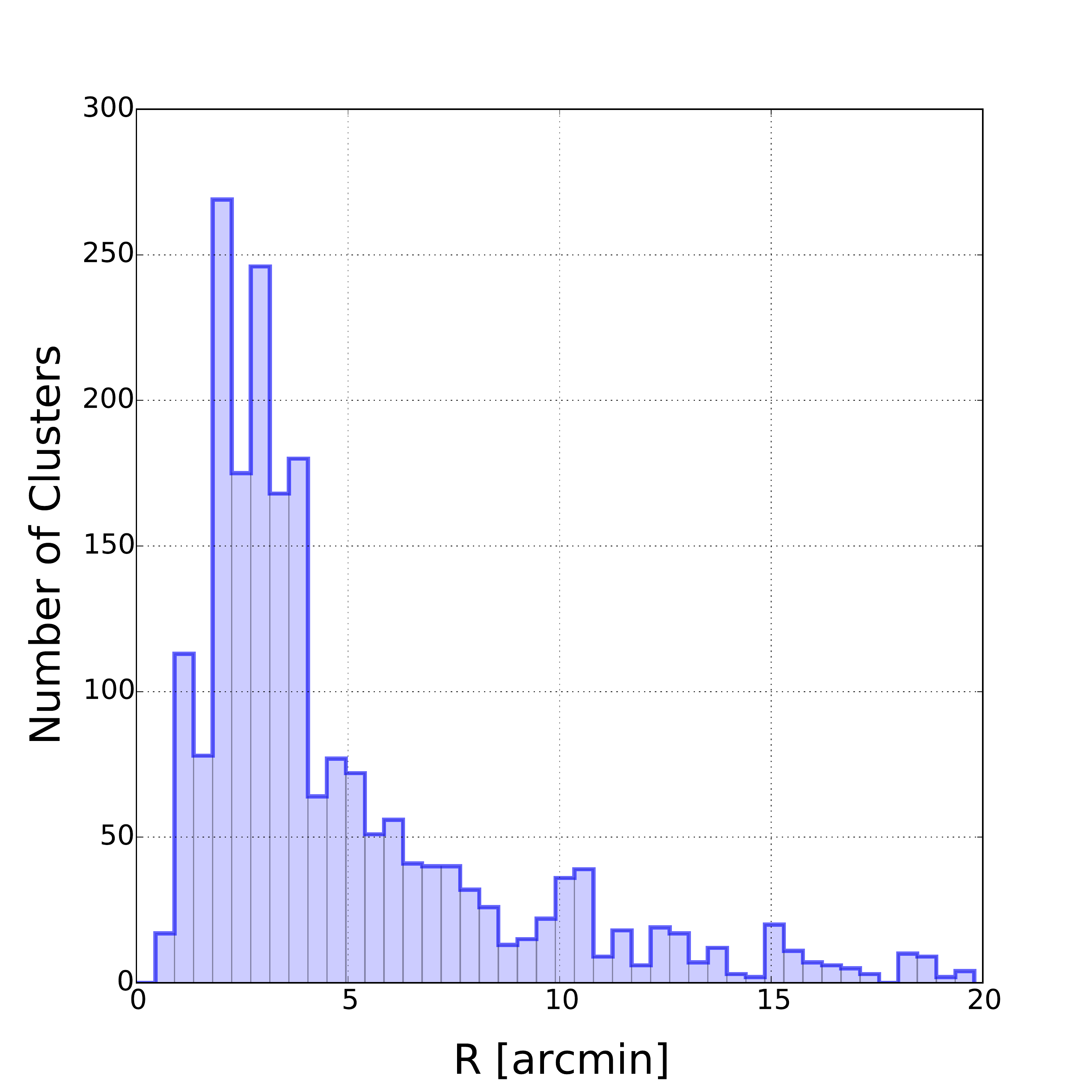} 
\includegraphics[width=8cm]{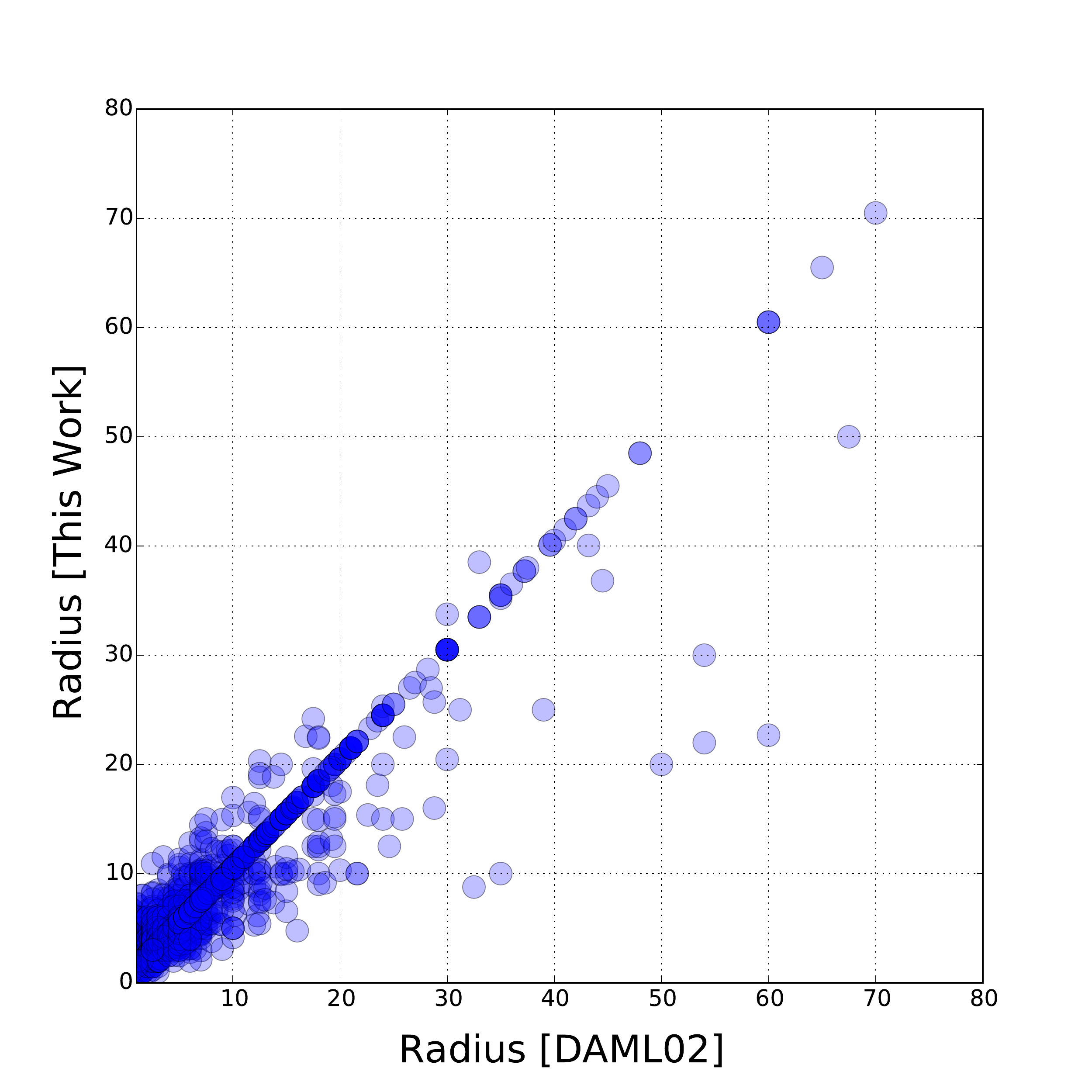} 
\caption{Top: The distribution of apparent angular cluster  radii estimated in this work up to 20 arcminutes. Bottom: a comparison between the radii estimated in this work and those from DAML02 (in arcminutes).}
\label{radius}
\end{figure}

Figure \ref{radius} shows the distribution of the apparent angular radii obtained in this work for the 2167 clusters listed in DAML02 up to an estimated radius of 20 arcminutes. It also compares our final cluster radii and those listed in DAML02. We consider that the observed differences are mainly due to the fact that DAML02 is a compilation of different works based on different datasets. In addition, we compare our final estimated radii with those from \citet[][hereafter K13]{2013A&A...558A..53K}. The authors derived cluster radii, among other parameters, for 3006 stellar clusters using the PPMXL \citep{2010AJ....139.2440R} and the 2MASS \citep{2006AJ....131.1163S} catalogues. In K13 the mode of the cluster radii distribution is about 7 arcminutes while the same central value  for our sample is about 2.5 arcminutes. The difference is quite big; however we should  stress again that we are deriving a functional radius from and for a single catalogue. To consider this estimate as the true radius of the cluster, if that concept existed and was unique and absolute, would be an error.  

However, going further into the understanding of this difference we can envisage several causes. The observed disagreement may be mainly due to the different techniques and different catalogues used to estimate the cluster radii. We have estimated apparent angular radii from all stars in the field while K13 obtains the radii values from member stars, using a quite different  procedure \citep{2005A&A...438.1163K, 2012A&A...543A.156K}. Therefore, the direct comparison must be taken with caution since it may be biased showing a strong disagreement. In addition, although \citet{2005A&A...438.1163K} and K13 obtain similar cluster radii, several studies \citep{2006AJ....132.1669S, 2011AcA....61..231B} have claimed  that using different data sources may lead to different cluster radius estimations.

We have also used the \textit{Digitized Sky Survey} (DSS)\footnote{http://archive.stsci.edu/dss/} to re-determine the central positions, of 10 open clusters, according to our best analysis of the RDPs. The new central coordinates were estimated as the positions maximizing the central star density. Table \ref{table1} includes the new coordinates as  well as those listed in DAML02 for the ten clusters  that showed quite  different  central positions.  
\begin{table}
	\centering
	\caption{\small{Redetermined central coordinates (J2000.0) for ten open clusters are shown  in columns 2 and 3. The coordinates from DAML02 are also included in  columns 4 and 5 for comparison.}}
        \vspace{0.2cm}
        \scalebox{0.9}[0.9]{
	\label{table1}
	\begin{tabular}{lcccc} 
		\hline
		\textbf{Name}    &$\alpha_{This work}$&$\delta_{this work}$    &$\alpha_{DAML02}$&$\delta_{DAML02}$ \\ 
                                &          (h:m:s)        &           (g:m:s)             &            (h:m:s)          &        (g:m:s)               \\ 
      \hline
Berkeley 28           &  06 52 07             &  02 54 47     & 06 52 12         &          02 56 00       \\ 
 Berkeley 39          &  07 46 48             & -04 40 06     & 07 46 42         &        - 04 36 00       \\ 
 Berkeley 43          &  19 15 32             &  11 16 20     & 19 15 36         &          11 13 00       \\ 
 Berkeley 45          &  19 19 05             &  15 42 47     & 19 19 12         &          15 43 00       \\ 
 Berkeley 50          &  20 10 01             &  34 57 58     & 20 10 24         &           34 58 00      \\ 
 BH 208               &  16 59 32             & -37 07 20     & 16 59 36         &          -37 05 00       \\ 
 IC 1311              &  20 10 46             &  41 10 27     & 20 10 18         &           41 13 00      \\ 
 IC 1369              &  21 12 06             &  47 46 04     & 21 12 06         &           47 44 00      \\ 
 IC 361               &  04 18 54             &  58 15 00     & 04 19 00         &           58 18 00      \\ 
 Ruprecht 164         &  11 30 25             &  -60 45 10    & 11 30 51         &          -60 44 00       \\ 

		\hline
	\end{tabular}}
\end{table}

For each cluster, once the UCAC4 data was downloaded, we discarded stars without proper-motion data or with errors in the proper motions larger than 12 mas yr$^{-1}$. The latter stars should be considered as observational fails and they might not correspond to real stars according to \citet{2013AJ....145...44Z}. In addition, we exclude the clusters with fewer than 20 stars and more than 10$^{4}$ stars. Cluster regions with few stars may suffer from subsampling effects as a result of patchy and heavy absorptions or incompleteness due to the photometric depth of the catalogue. On the other hand, open clusters with the largest angular diameters show a low surface brightness with high contamination of fore/background stars. This effect may prevent a proper determination of the cluster members using these methods. This pruning  leads to a sample of 1876  open clusters analysed in this study.

%**********************************************************************
%**********************************************************************
\section{Methods used in the membership determination}
\label{meth}
%**********************************************************************
%**********************************************************************

In this work we apply three different methods to determine the membership for the previously selected 1876 open clusters. The first method (hereafter M1) is fully described in \citet{2016MNRAS.457.3949S}. Basically, it uses different sets of variables satisfying the simple condition of being more densely distributed for the cluster members than for the field stars. The membership probabilities are estimated in a 1-Dimensional space, defined by the Euclidean distances between every star and the cluster central overdensity, in the variables space of N-Dimensions. Therefore, it reduces the estimation of membership probabilities from an N-Dimensional space to a 1-Dimensional one. The method involves two iterative processes. Initially, the distances between every star and the cluster centre are estimated. Then, the distance distribution of the stars is modeled by a mixture of two 1-Dimensional Gaussians, one for the cluster members and other for the field stars. The parameters defining the total model and the membership probabilities are calculated through an iterative Wolfe estimation procedure \citep{Wolfe}. The stars with a membership probability higher than 0.5 are selected as members according to the Bayes minimum error rate decision rule \citep{bayes_rule}. In this work, this method is applied to the proper motion data.

The second method (hereafter M2) is a Bayesian, non-parametric one developed by \citet{1990A&A...235...94C}. It determines the members of the clusters using the positions (angular distances) and the proper motion data without any a priori assumptions about the cluster and field star distributions, but  assuming two hypotheses: i) there are two populations in the field; cluster members and field stars, and ii) the cluster members are more densely distributed in the phase space. Membership probabilities are calculated by using Gaussian Kernel estimators in an iterative way through a discriminant analysis. In every iteration, three different probabilities for each star are estimated: one just using the positions of the stars, another using only the proper-motion data (kinematic probability), and the last using both positions and proper motions (joint probability). As a first option, cluster members in every iteration are selected as those stars with joint and kinematic probabilities higher or equal to 0.5. 

The third method (hereafter M3) follows a parametric approach, fully described in \cite{1985A&A...150..298C}. It uses only the stellar proper motions  to determine the membership probabilities. This method fits the PDF of the whole sample by a mixture of two bivariate Gaussian distributions, one for the cluster members and other for the field stars. Through an iterative Wolfe estimation procedure, the parameters that defines the total model are calculated as well as the membership probabilities. The cluster members are those stars with a membership probability higher than 0.5. 

Before starting with the membership analysis, we detect and remove sample outliers. In this task, we use a non-parametric technique \citep{1985A&A...150..298C}, which estimates the probabilities of the stars being outliers of the parent sample using the proper-motion data. In this way, objects with a probability of being an outlier greater than 0.5 were rejected for further analysis.

%**********************************************************************
%**********************************************************************
\section{Results}
\label{results}
%**********************************************************************
%**********************************************************************

In this section, we compare the level of agreement achieved by the different methods described in Section \ref{meth}. Furthermore, we compare our results with previous catalogues that also contain the main physical variables for each cluster, such as: DAML02, K13 and D14. The large number of objects listed in these catalogues will enable a comparison with a reliable statistical  significance. We use the estimated probabilities, derived from the three different methods (the joint probability in the case of M2) as weight factors to calculate the cluster mean proper motions $\mu_{\alpha}cos\delta$ and $\mu_{\delta}$, as:  

\begin{equation*}
   \mu_{\alpha}cos\delta = \frac{\sum\nolimits_{i=1}^{n}P_i \mu_{\alpha}cos\delta_i}{\sum\nolimits_{i=1}^{n} P_i}
\end{equation*} 

\begin{equation*}
\mu_{\delta} = \frac{\sum\nolimits_{i=1}^{n}P_i \mu_{\delta, i}}{\sum\nolimits_{i=1}^{n} P_i}
\end{equation*}

\noindent where $\mu_{\alpha}cos\delta_i$ and $\mu_{\delta, i}$ are the proper motion and P$_i$ is the probability estimated by each method for the $\textit{i}$-th star. Similarly, the cluster proper-motion dispersion and the correlation coefficients have also been estimated. The database containing the results is available in electronic format at the SSG-IAA\footnote{http://ssg.iaa.es/en/content/sampedro-cluster-catalog} website as two complementary files. The main parameters obtained by the three methods are included in a general catalogue. We also generate individual cluster-by-cluster files giving the membership probabilities and additional information from the UCAC4 catalogue (more info in Appendix \ref{cat}).

%**********************************************************************
\subsection{Comparison of the results obtained with the three different methods}
\label{comparison}
%**********************************************************************

In this study, we investigate 1876 open clusters using the three different methods described in Section \ref{meth}. Whereas M1 converges to a solution for a total of 1748 clusters (93\%), M2 and M3 do so for a total of 1693 (90\%) and 1819 (97\%), respectively.  The three different approaches converge for 1584 (84\%) clusters in common. 

\begin{figure*}
\includegraphics[width=5.8cm]{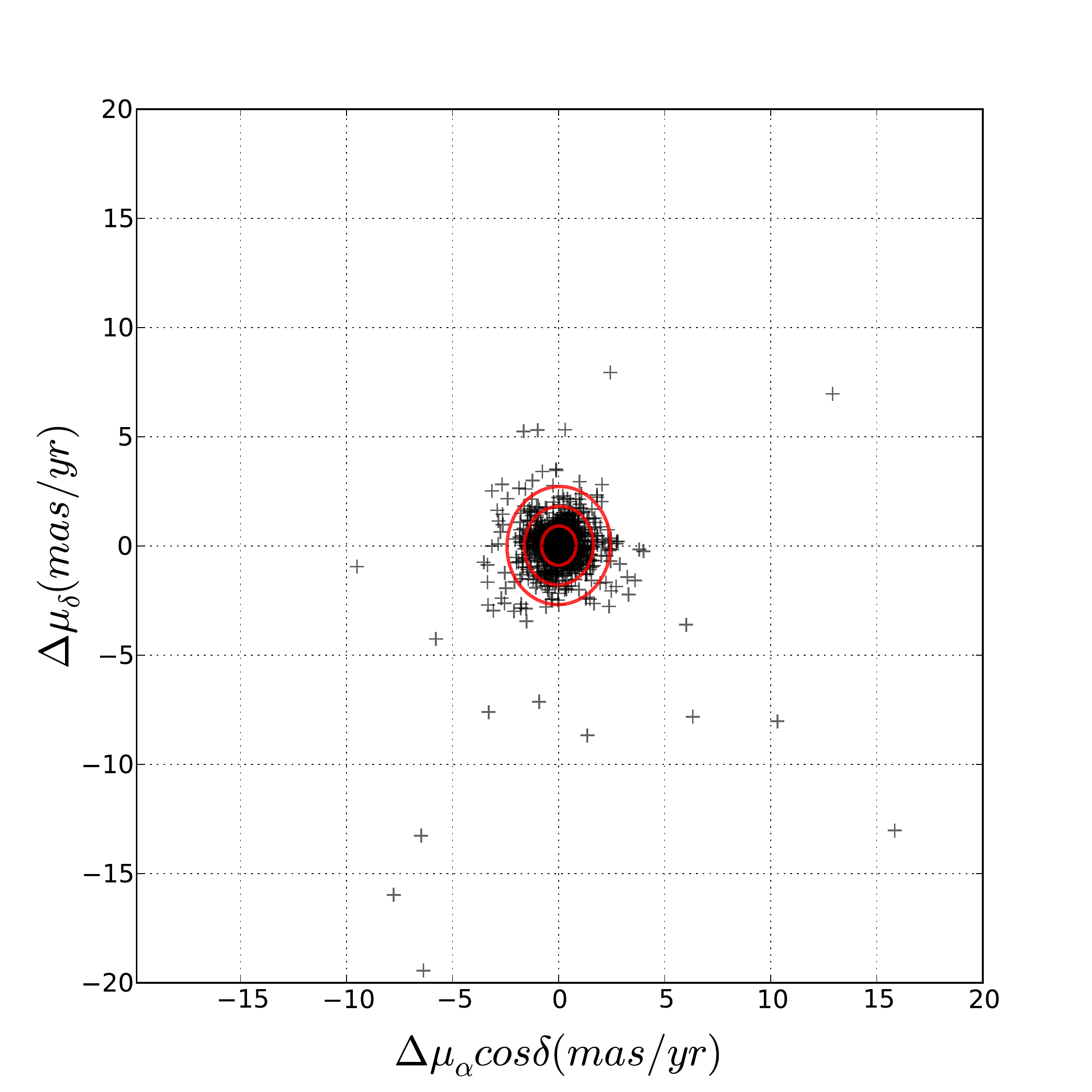} 
\includegraphics[width=5.8cm]{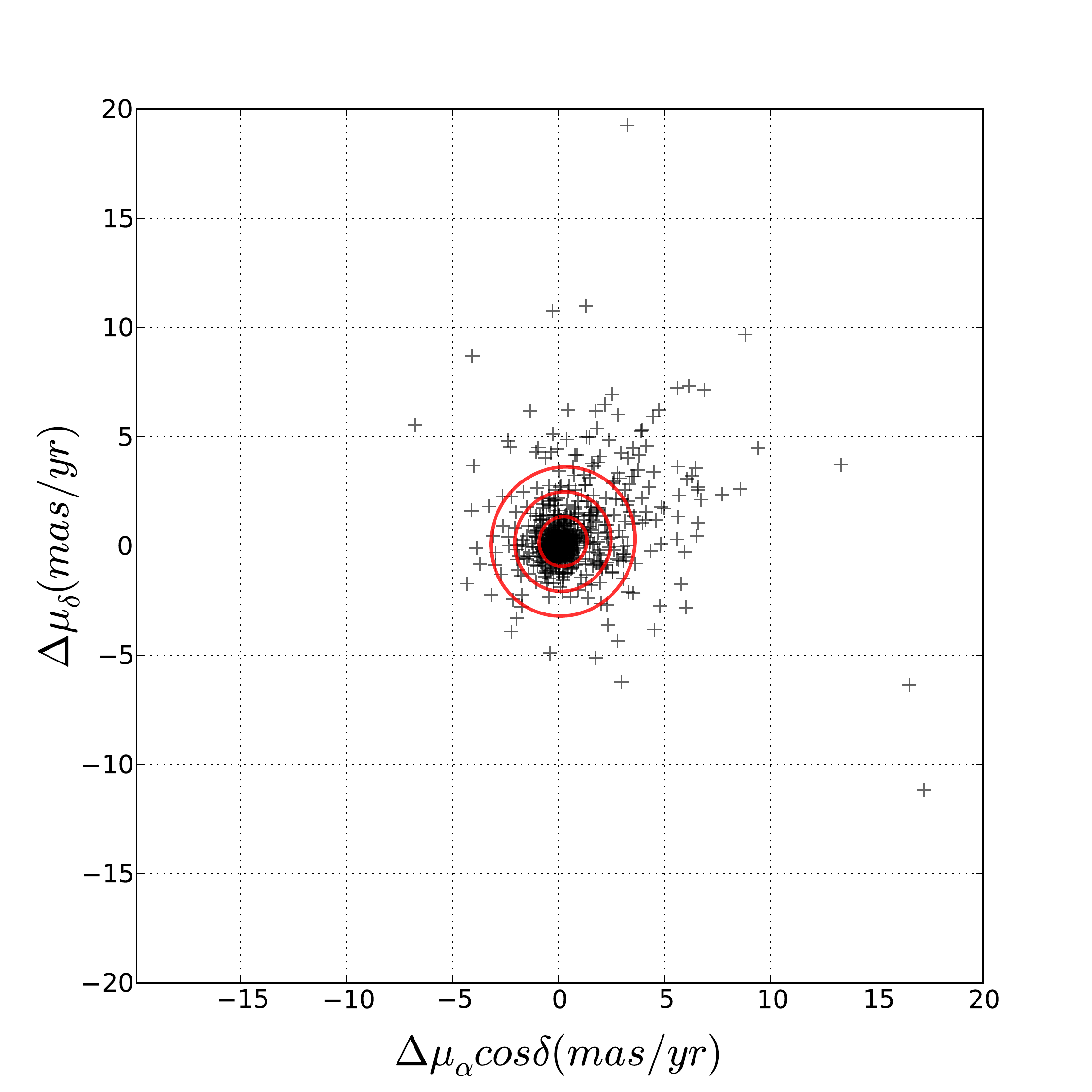} 
\includegraphics[width=5.8cm]{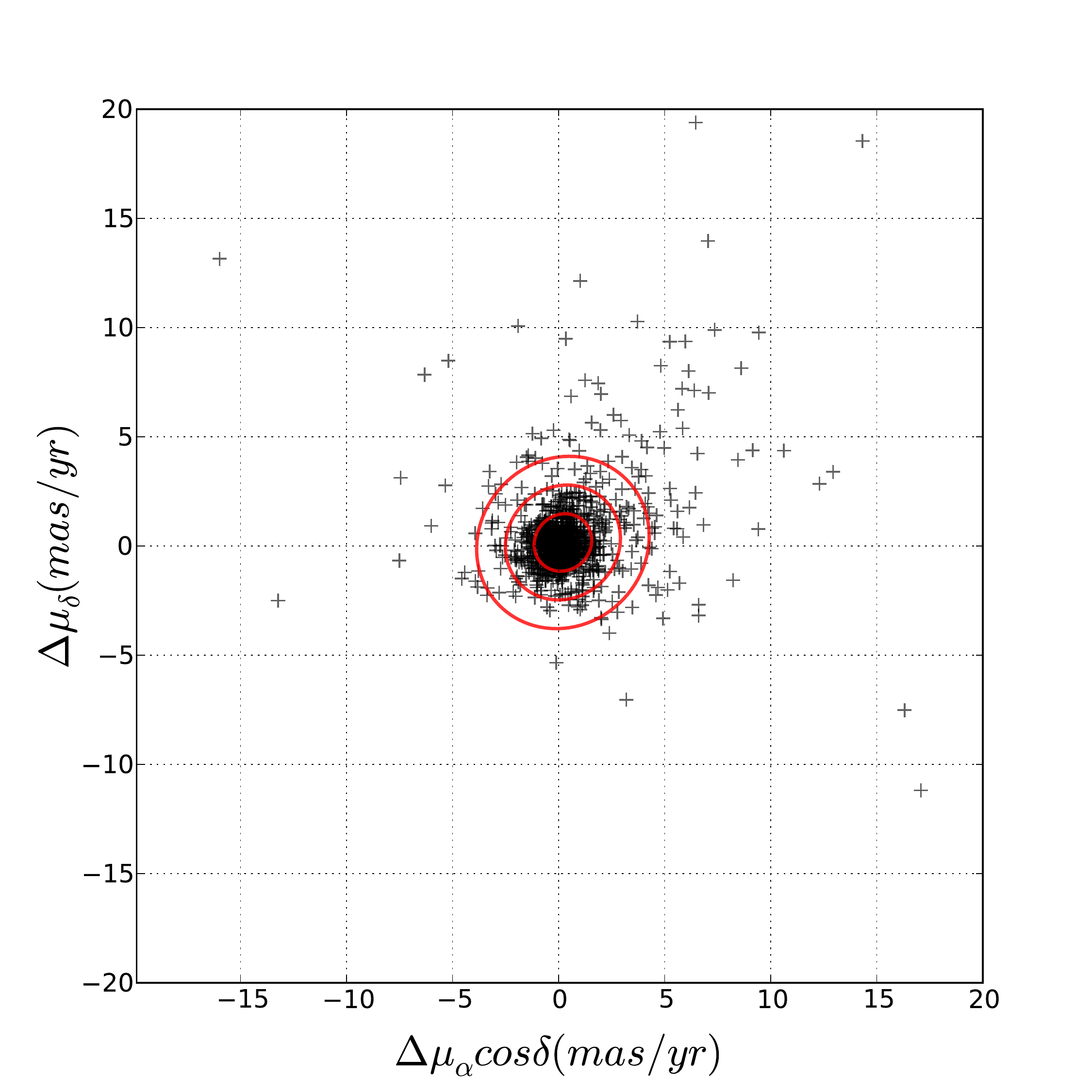} 
\caption{Left: Differences in the values of the clustr mean  proper motions between M1 and M2. Centre: Differences in the values of the cluster mean proper motions between  M1 and M3. Right: Differences in the values of the cluster mean proper motions between  M2 and M3. The red ellipses represent the 1, 2 and 3$\sigma$ dispersion levels, respectively.} 
\label{comp_todas}
\end{figure*}

Table \ref{table2} and Figure \ref{comp_todas} show the results of the comparison in $\mu_{\alpha}cos\delta$ and $\mu_{\delta}$ between the methods. In Table \ref{table2}, $\Delta\mu_{\alpha}cos\delta$ and $\Delta\mu_{\delta}$ are the differences in the values of the cluster mean proper motions, $\sigma_{\Delta\mu_{\alpha}cos\delta}$ and $\sigma_{\Delta\mu_{\delta}}$ represent their dispersions and $N$ the number of common clusters. The values of the mean differences are close to 0 mas yr$^{-1}$ with small dispersion (lower then 1.4 mas yr$^{-1}$), indicating good agreement between the methods. Furthermore, the mean differences among the methods, for the second moments of the cluster member distribution, are lower than the typical proper-motion errors in the UCAC4 catalogue, overall when comparing  M1 and M3. The agreement also occurs in the number of cluster members estimated. However, for some cases it is very difficult to distinguish between members and background stars, especially in regions associated to a patched absorption pattern or to clusters with large apparent angular radii, i.e., with a low contrast between the cluster and the field. For these cases, we notice that M2 tends to determine lower fractions of members than M1 and M3. It is worth noting that M2 provides a different cluster view than that derived from the M1 and M3 methods; M2 makes uses of four astrometric variables, while M1 and M3 do so for the two proper-motion components. Nevertheless, in this scenario, the results obtained from all methods should be considered with caution. 

Finally, for six clusters (Alessi 53, Berkeley 76, Czernik 11, Loden 27, Pismis 5 and Ruprecht 3) none of the methods were capable of converging  to a solution. This is not surprising since these clusters have few stars, with sparse distributions in the position and in the proper-motion spaces, showing the typical problems derived from small-number statistics. Moreover, Loden 27 is flagged as a dubious open cluster in DAML02 and Ruprecht 3 is not even considered as a physical system by \citet{2017MNRAS.466..392P}.

\begin{table}
	\centering
	\caption{Comparison of the cluster mean proper motions obtained by the methods (see the text for details). $\Delta\mu_{\alpha}cos\delta$, $\Delta\mu_{\delta}$, $\sigma_{\Delta\mu_{\alpha}cos\delta}$ and $\sigma_{\Delta\mu_{\delta}}$ are expressed in mas yr$^{-1}$. $N$ is the number of clusters for which the compared methods converge to a solution.}
       \vspace{0.2cm}
       \scalebox{1.1}[1.1]{
	\label{table2}
	\begin{tabular}{lccc} 
		\hline
              \textbf{Methods}                             &    \textbf{M1-M2}    &    \textbf{M1-M3}     &     \textbf{M2-M3}  \\ 
		\hline
              $\Delta\mu_{\alpha}cos\delta$                &	       0.01          &	      0.21               &	             0.20                  \\
              $\Delta\mu_{\delta}$                               &	       0.02          &	      0.20               &	             0.16                   \\
              1$\sigma_{\Delta\mu_{\alpha}cos\delta}$ &	       0.81           &	      1.11               &	             1.39                   \\
              1$\sigma_{\Delta\mu_{\delta}}$                 &	       0.90           &	      1.16               &	             1.28                    \\
              $N$                                                          &	       1606          &	      1713              &	             1661                   \\
		\hline
	\end{tabular}}
\end{table}

%**********************************************************************
\subsection{Comparison with other catalogues}
\label{comp_lit}
%**********************************************************************

A direct comparison of the results presented in this work with those in the literature is not straightforward, since they use different data, methods and criteria to select  members, which may lead to different samples of stars to determine the mean proper motions of the clusters. Therefore, we opt to compare our results with those published catalogues that present a large number of open clusters, such as DAML02, K13 and D14. 

The DAML02 catalogue is a compendium of results published from different authors, thus it is based on different data and methods but presents the mean proper motions in the Hipparcos System and has been widely used in recent years. In K13 the cluster membership,  and mean proper motions are estimated for more than 3000 open clusters, using the PPMXL and the 2MASS catalogues as the main observational basis. The authors use an interactive human control of a standardized set of multi-dimensional diagrams to determine kinematic and photometric membership probabilities for stars in the cluster region. D14 is a homogeneous study of 1805 open clusters using the UCAC4 catalogue. The authors determine mean proper motions and membership probabilities using a parametric model, fitting the PDF for the whole sample by two bivariate Gaussian functions, also taking  into account the proper-motion errors of the stars.

Table \ref{table3} and Figure \ref{comp_lit} show the results of the comparison in $\mu_{\alpha \cos{\delta}}$ and $\mu_{\delta}$ for each cluster, reproducing very well, on average, the previous determinations. Dispersion differences lie below 2.8 mas yr$^{-1}$ in the comparison with D14 and DAML02 for the three methods. This indicates that  there is not  a statistically significant separation between the compared  distributions,  taking into account the catalogued proper-motion  errors. Basically, the higher dispersion in the comparison with DAML02 is likely due to the heterogeneity of the proper-motion sources used in DAML02.  A few open clusters in Figure \ref{comp_lit} present differences greater than the 3$\sigma$ level. Generally after a visual inspection, we notice that most of these open clusters do not show clear over-density in the proper-motion space, using UCAC4 data. Likewise, some clusters such as Stock 1 and Alessi 13 present a  high degree of field contamination and in both cases a proper determination of the cluster members  is difficult . The greatest individual  differences are found when  comparing with K13, which is likely caused, as previously discussed,  by the different data sources and methods used for the cluster membership analysis. Again, we notice that most of these differences are related with clusters that do not show a clear over-density in the UCAC4 proper-motion space at the position indicated in K13.

We observe that D14 does not properly determine mean proper motions for the large and nearby clusters NGC 7092, Ruprecht 147, Mamajek 1, Blanco 1, NGC 752 and Stock 2. For these clusters, M1 and M3 obtain satisfactory results which agree with those published by \citet{2001A&A...376..441D} based on the Tycho-2 data \citep{2000A&A...355L..27H}. 

Finally, we determine, for the first time, membership and mean proper motions of 18 new cluster candidates. The candidates correspond to a cluster sample with apparent diameter smaller than 3 arcminutes in DAML02. For most cases, there has been no confirmation of the real existence of the clusters through photometric analysis and the new cluster membership study could help to verify this issue.

\begin{table*}
	\centering
	\caption{Comparison of the mean proper motions obtained by the three methods used in this work with those published by DAML02, K13 and D14.  $\Delta\mu_{\alpha}cos\delta$, $\Delta\mu_{\delta}$, $\sigma_{\Delta\mu_{\alpha}cos\delta}$ and $\sigma_{\Delta\mu_{\delta}}$ are expressed in mas yr$^{-1}$. $N$ is the number of clusters in common between each method and the catalogue indicated.}
       \vspace{0.2cm}
       \scalebox{0.95}[1.]{
	\label{table3}
	\begin{tabular}{lccccccccc}
		\hline
              \textbf                                       &    
\textbf{D14-M1} & \textbf{D14-M2} & \textbf{D14-M3} &    \textbf{K13-M1} 
& \textbf{K13-M2} & \textbf{K13-M3} & \textbf{DAML02-M1} & 
\textbf{DAML02-M2} & \textbf{DAML02-M3}  \\
		\hline
              $\Delta\mu_{\alpha}cos\delta$ &	      0.03              &	  
        0.04       &        -0.20           &	    0.80               &	 
0.82            &      0.46        &	       0.19                 &	      
        0.21           &	               -0.10              \\
              $\Delta\mu_{\delta}$                &          0.01         
     &		 0.00       &	-0.21           &	    -0.50              &	 -0.44   
        &      -0.80        &	      -0.13                 &	            
-0.08           &	               -0.37              \\
              $\sigma_{\Delta\mu_{\alpha}cos\delta}$ & 0.80       &	      
    1.01        &	 1.35           &	    3.00               &	  3.08       
     &       3.29        &	       1.89                  &	     2.03       
     &	        2.33             \\
              $\sigma_{\Delta\mu_{\delta}}$  &	      0.74              &	 
          0.91       &	  1.36          &	    3.15               &	   3.31 
          &        3.43       &	       1.79                   &	     1.92 
           &	        2.78             \\
              $N$                                           &         
1599             &	         1558       &	 1648          &	    1563       
        &	    1513        &        1625      &	       1733                
   &	     1675           &	        1800             \\
		\hline
	\end{tabular}}
\end{table*}

\begin{figure*}
\includegraphics[width=5.8cm]{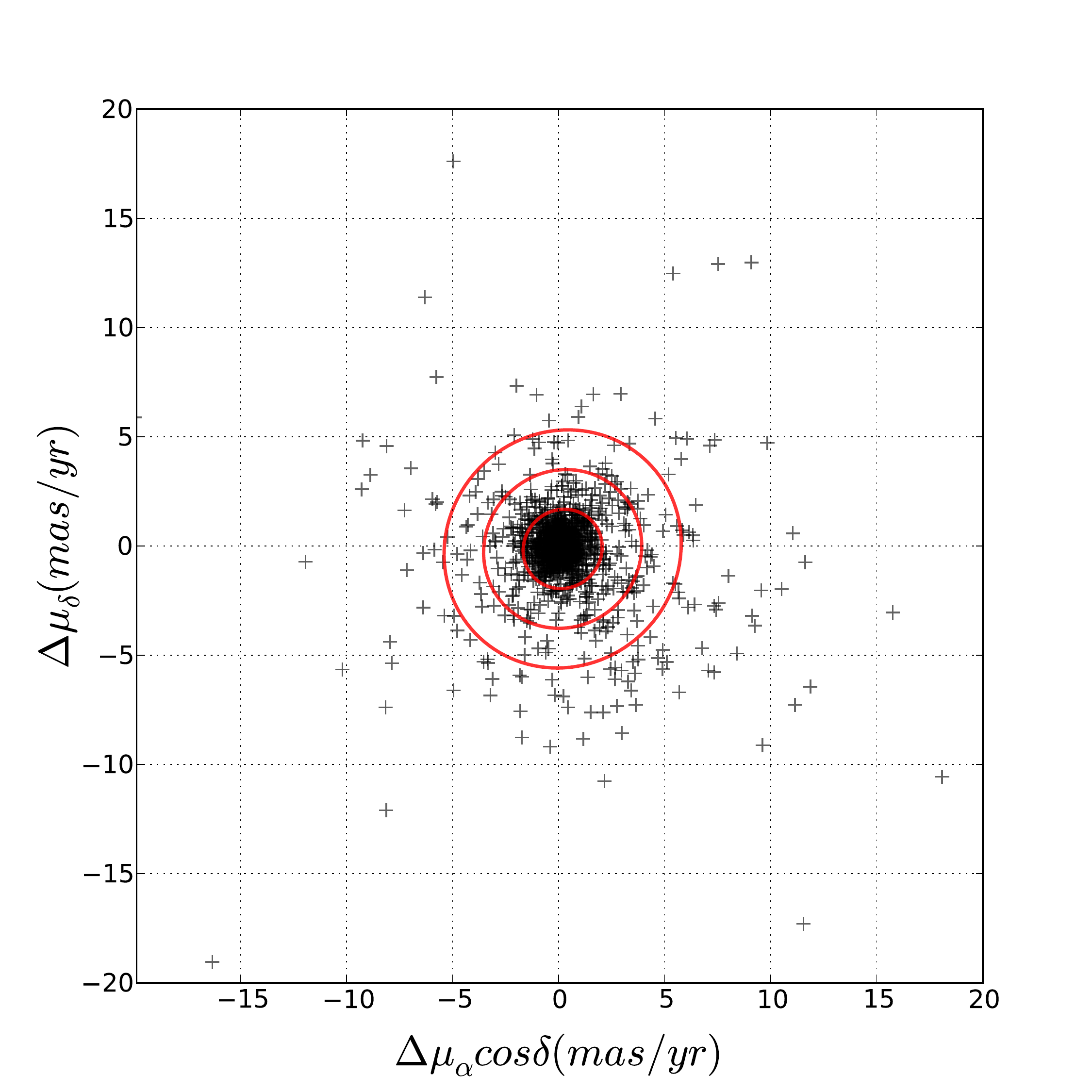} 
\includegraphics[width=5.8cm]{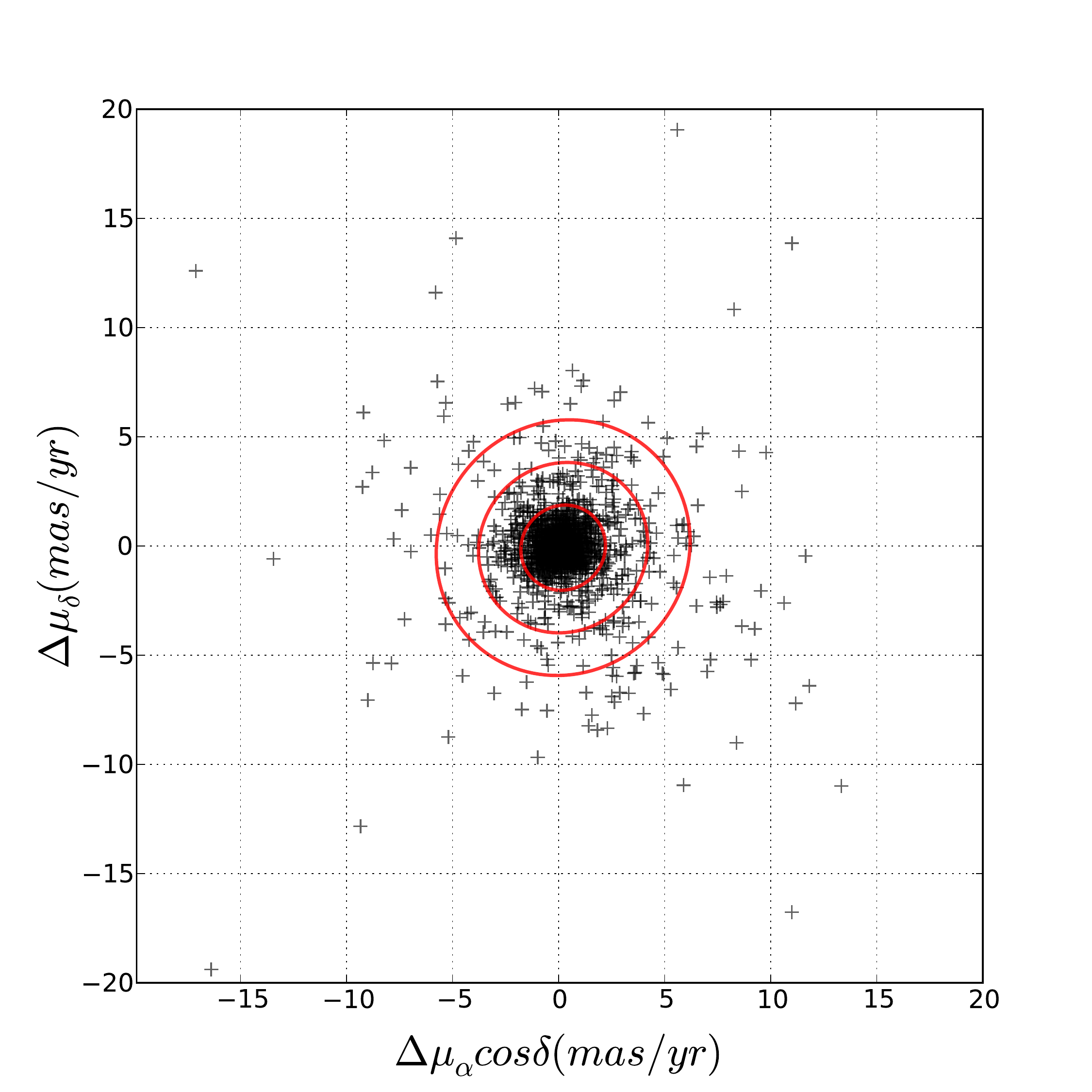} 
\includegraphics[width=5.8cm]{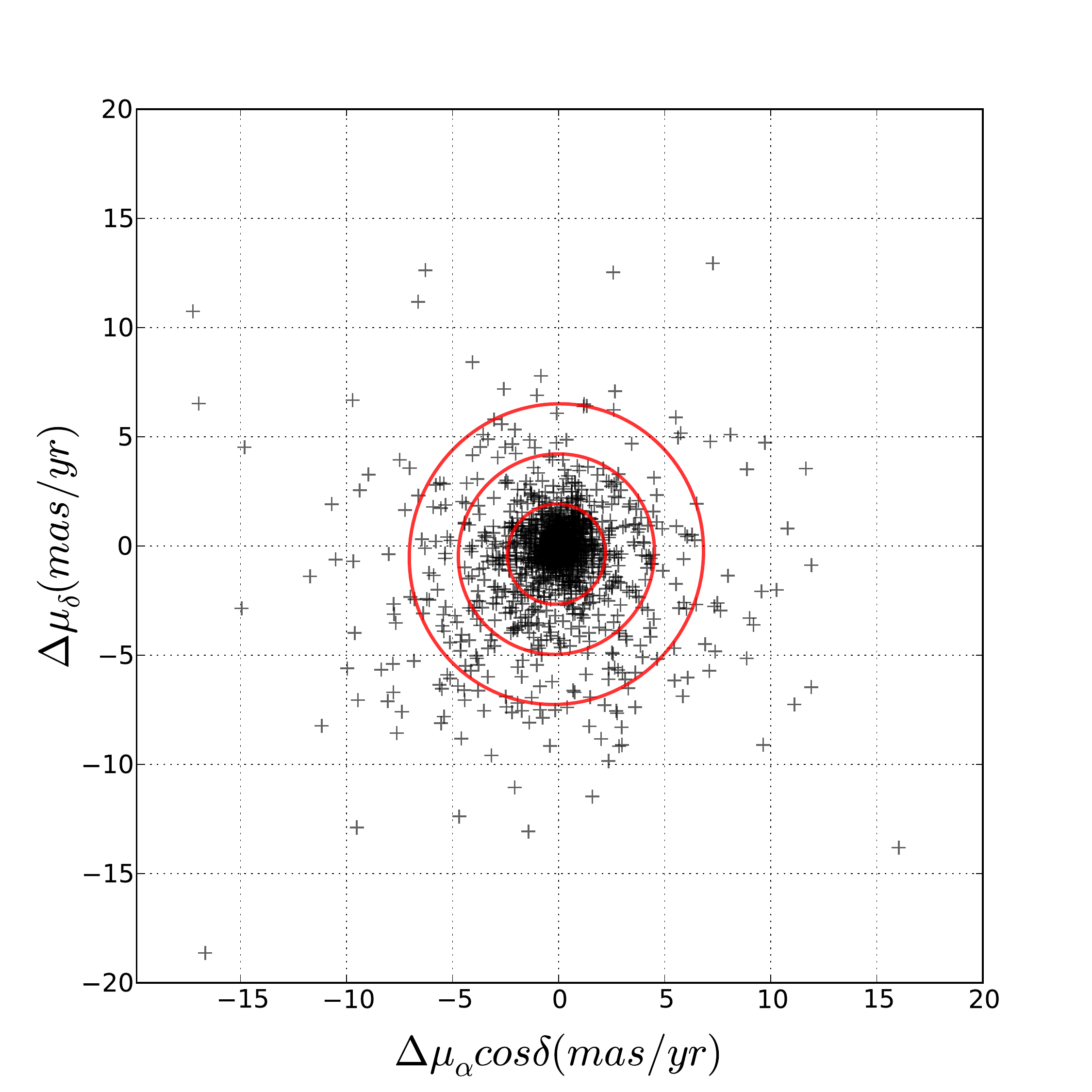} 
\includegraphics[width=5.8cm]{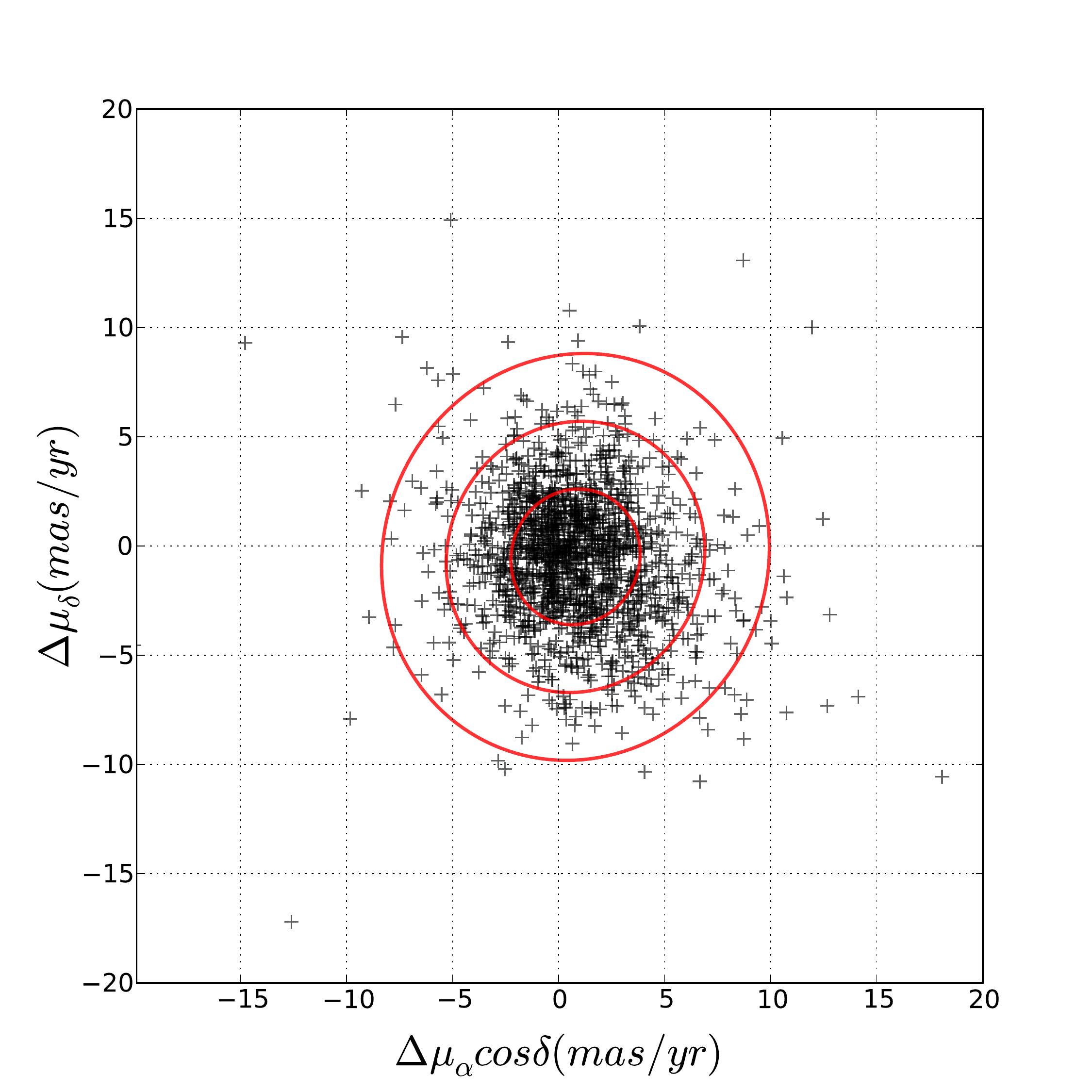} 
\includegraphics[width=5.8cm]{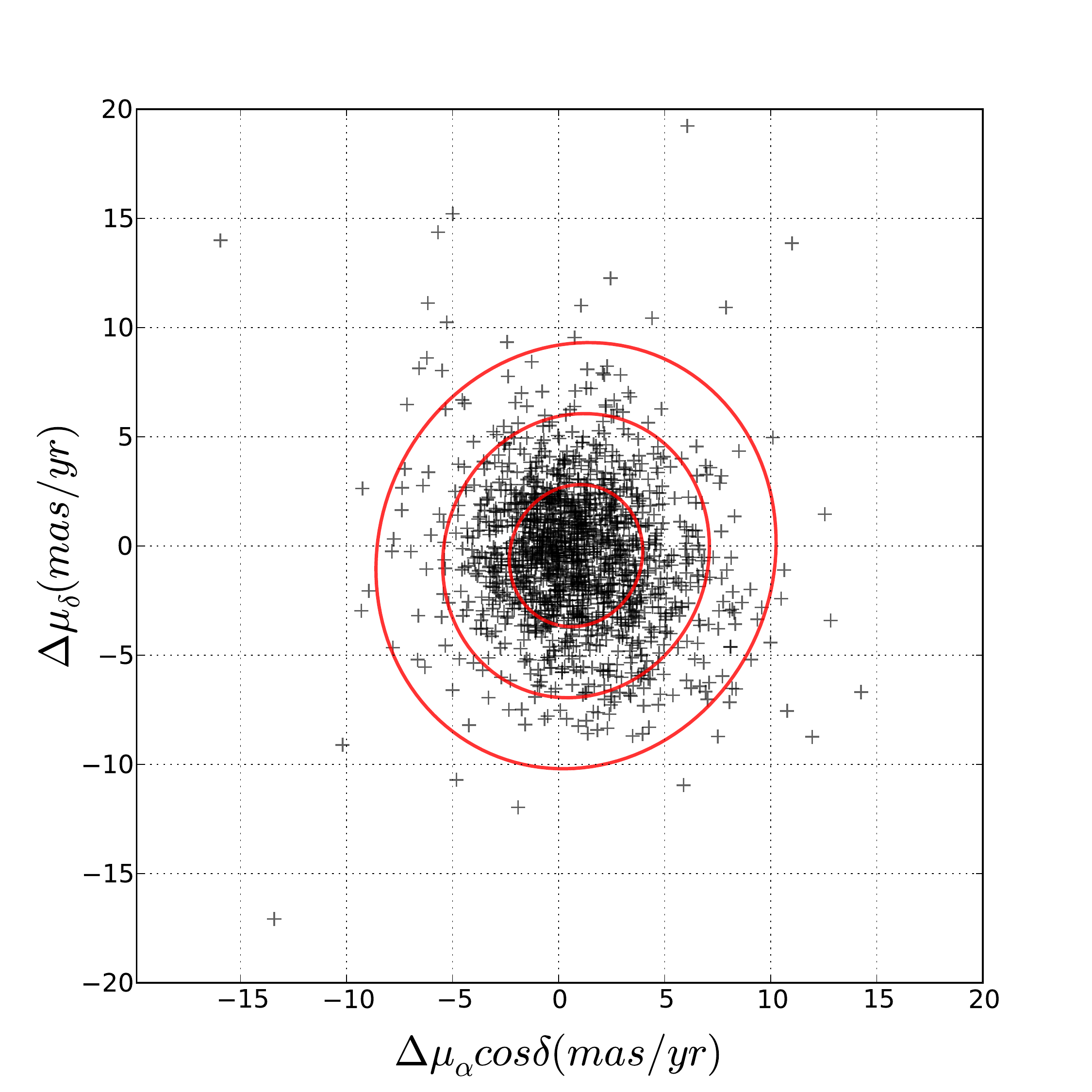} 
\includegraphics[width=5.8cm]{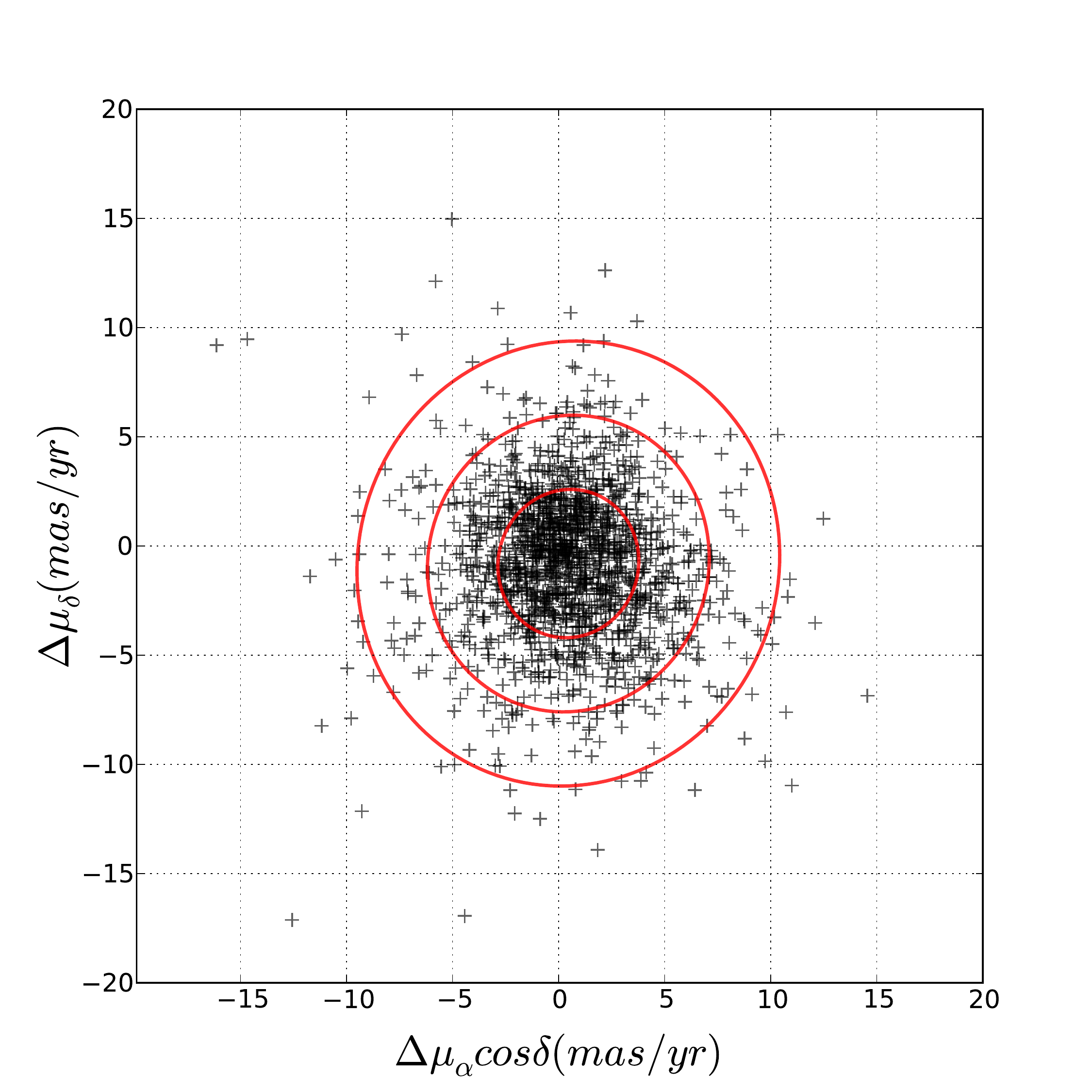}\\ 
\includegraphics[width=5.8cm]{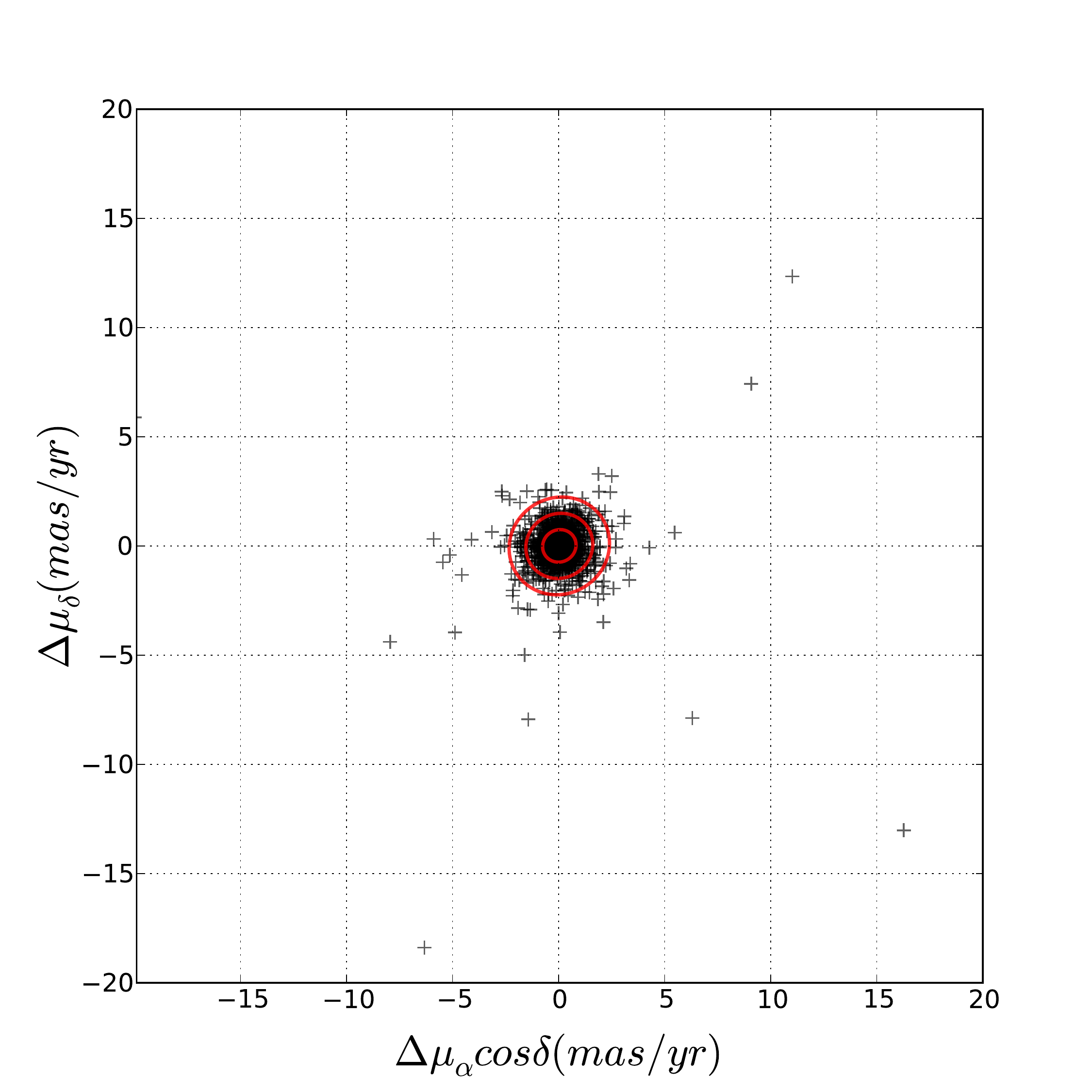} 
\includegraphics[width=5.8cm]{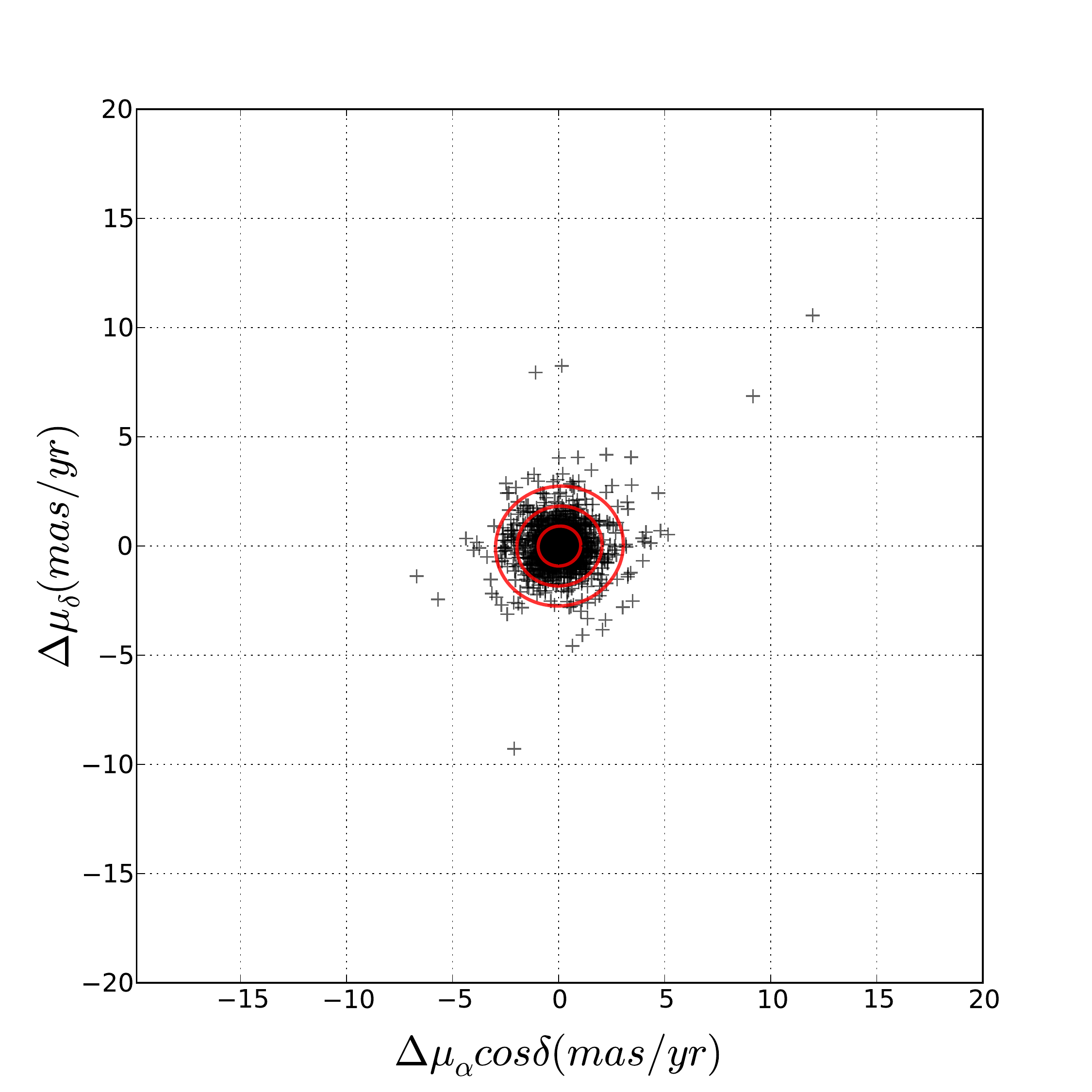} 
\includegraphics[width=5.8cm]{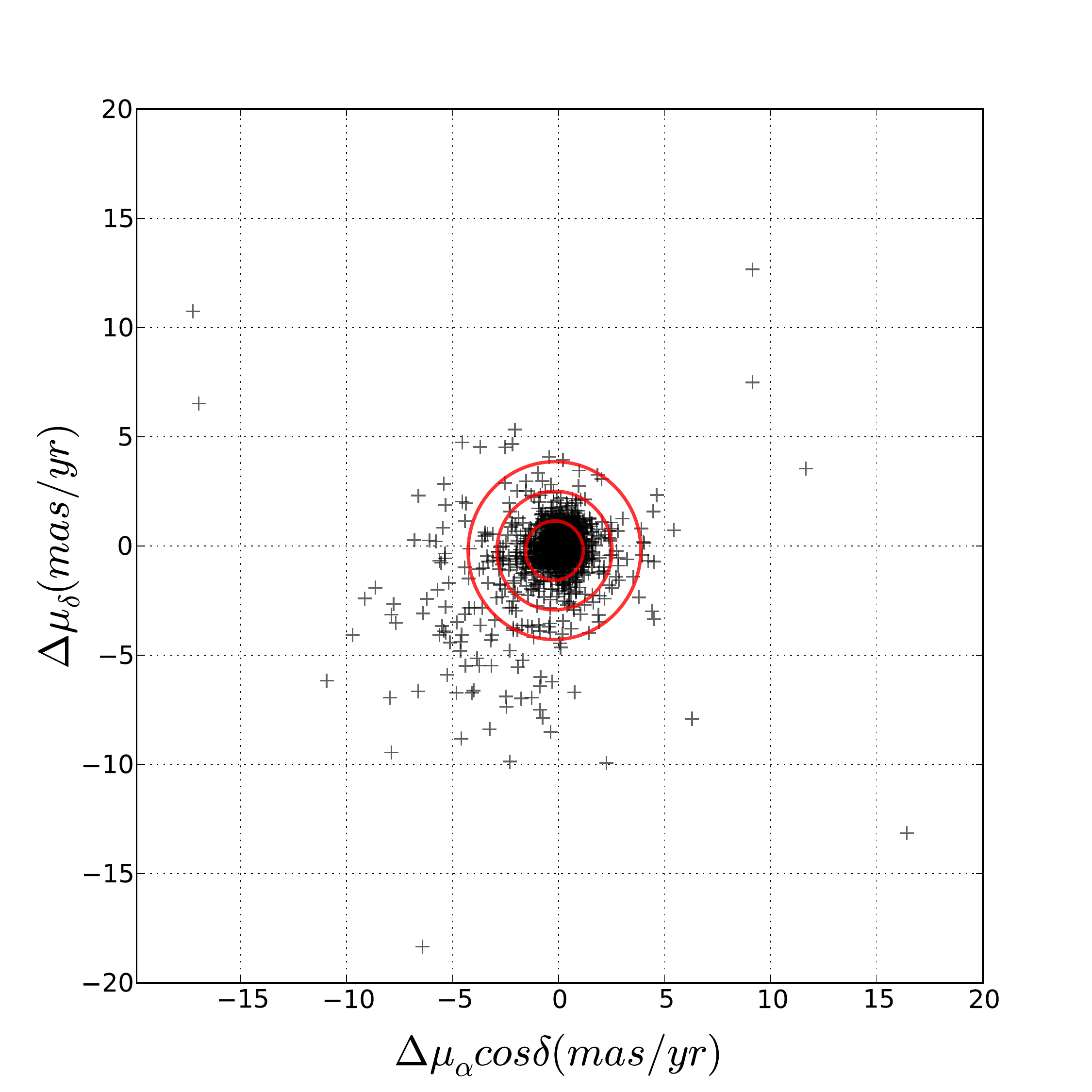}\\ 
\caption{Comparison of our results for mean proper motions with those published in the literature. The first line presents the comparison with DAML02, the second line with K13 and the last line with D14. From left to right the plots present the comparison with M1, M2 and M3 methods, respectively. The ellipses represent the 1, 2 and 3$\sigma$ dispersion levels and their values are given in Table \ref{table3}.}
\label{comp_lit}
\end{figure*}

\begin{table*}
\caption{Mean proper motions and dispersions, expressed in mas yr$^{-1}$, for 18 open clusters candidates with unpublished data in DAML02.}
\label{ineditos}
\vspace{0.5cm}
\begin{centering}
%\begin{turn}{90}
\scalebox{1.}[1.]{
\begin{tabular}{lcccccc}
     \hline
        \textbf{Cluster Name}          &	   \multicolumn{2}{c}{M1}             &	   \multicolumn{2}{c}{M2}              &	   \multicolumn{2}{c}{M3}            \\ 
     \hline                                                                        
	                               &		   $\mu_{\alpha} cos\delta$, $\sigma_{\mu_{\alpha} cos\delta}$  &  $\mu_{\delta}$, $\sigma_{\mu_{\delta}}$    &	   $\mu_{\alpha} cos\delta$, $\sigma_{\mu_{\alpha} cos\delta}$  &  $\mu_{\delta}$, $\sigma_{\mu_{\delta}}$    &	   $\mu_{\alpha} cos\delta$, $\sigma_{\mu_{\alpha} cos\delta}$  &  $\mu_{\delta}$, $\sigma_{\mu_{\delta}}$      \\                 
     \hline
     AH03 J1725 34.4     &	-3.40, 1.26    &	0.74, 1.05      &	         -4.06, 5.40   &	-0.33, 4.53      &	 -3.46, 1.24     &  0.79, 1.01    \\    
     Alessi 52                  &	-5.78, 5.83    &	0.08, 4.54      &	         -5.58, 5.07   &	0.06, 4.27       &	 -5.86, 5.76     &  0.15, 4.43      \\
     BH 208                    &	      --             &	    --                &	         -5.25, 8.11    &	 -1.43, 7.60     &	 -8.44, 1.85   &	-2.37, 6.59     \\
     DC 1                        &	0.76, 1.68      &	 -0.34, 1.71     &	  0.90, 4.54    &	 -0.98, 4.29     &	 0.36, 1.96    &	-0.54, 1.99    \\
     Dutra Bica 12          &	1.27, 4.72      &	 -3.82, 5.07     &	 0.76, 7.78    &	 -3.66, 8.46     &	 1.23, 4.72    &	-3.81, 5.29       \\   
     FSR 0647                &	-2.64, 2.39     &	 0.70, 3.95      &	 -0.77, 13.53 &	 -1.95, 7.70     &	 -1.90, 6.61   &	0.36, 5.24    \\
     FSR 0696                &	       --             &	     --                &	 -7.27, 4.39   &	 -3.14, 3,39     &	        --           &	        --          \\
     FSR 0763               &	       --             &	     --                &        -12.68, 6.52  &   -4.59, 4.79     &	 -9.33, 2.10   &	 -2.67, 0.78  \\
     FSR 0814               &	0.42, 2.26       & -3.36, 2.04     &	0.68, 4.49      &	  -3.40, 4.13     &	 -0.90, 1.38   &	-5.39, 0.52   \\
     FSR 0828               &   0.97, 3.53       &	 -3.49, 3.29     &	 1.19, 4.83     &	  -3.39, 5.07     &	 0.96, 3.36    &	-3.50, 3.52  \\
     FSR 1308               &     --                   &	     --        &	        --           &	     --                 &	 -3.62, 2.01    &	1.04, 0.54\\        
     FSR 1343               &    --                    &     --                &	 -3.30, 4.42    &  3.84, 3.96       &	 -10.37, 0.45 &	-3.16, 2.08 \\
     Juchert 10              &	-4.65, 4.59      &  -5.37, 4.18     &	 -4.28, 5.00    &	  -4.99, 5.13      &	 -4.89, 2.05   &	-5.44, 3.51  \\
     Kronberger 23        &	-0.92, 3.96      &  -0.44, 2.84     &	  -0.78, 5.74   &	  0.20, 5.33       &	 -1.00, 4.11   &	-0.55, 2.79   \\
     Majaess 30             &	0.31, 3.52       &  -3.27, 3.28     &	  -1.38, 4.90   &	  -2.60, 4.28       &	  -0.71, 2.26  &	-3.72, 2.20  \\
     Majaess 9               &	-2.32, 4.66      &  2.51, 3.53     &	 -2.49, 3.73    &	  1.63, 2.94         &	  -2.32, 4.09  &	2.54, 3.55  \\
     Teutsch 127            &	-1.44, 2.29      &  -2.86, 1.60     &	 -0.26, 4.89    &	  -3.70, 4.79       &	  -3.39, 4.40  &	-1.99, 0.52  \\
     Wit 3                       &	0.18, 3.40       &   -3.89, 5.08     &	 1.32, 7.17     &	  -3.81, 5.41       &	  0.16, 1.54   &	-2.61, 5.71 \\
     \hline

\end{tabular}} 
%\end{turn}
\end{centering}
\end{table*}

%**********************************************************************
%**********************************************************************
\section{Conclusions}
\label{intro}
%**********************************************************************
%**********************************************************************

In this paper, we present a homogeneous multi-membership and mean proper motion catalogue of a sample of 1876 open clusters. The fundamental idea underlying this work is the comparison of different cluster membership analysis methods applied to the same set of data. In particular, we wished to thoroughly check with real data the method developed by \citet{2016MNRAS.457.3949S} (M1), which enables the utilization of multiple physical variables for cluster membership analysis. The increase in the number of variables in a statistical analysis entails the problem that for the same sample the volumetric density in the N-variable space decreases drastically as we increase the number of variables, making, in many cases, any later analysis inviable. In \citet{2016MNRAS.457.3949S} (M1) we avoided this problem by transforming the N-dimension space to a monodimensional one defined by the distance to the distribution centre in that N-variable space. This method was specially designed for the analysis of Gaia data, in which we would have complete information of the phase space for many clusters.  

The three different methods used in this study are based on the formulation and analysis of the distribution model of two populations in subspaces of phase space. The model chosen and the way in which the subsequent analysis is carried out can represent different aspects of what we could call a stellar cluster. In most cases, if the astrometric data distributions are well behaved, the results derived from the three analyses coincide. However, clear differences can appear if the spatial and kinematic (proper motions) structure of the cluster deviates from Vasilevskis' model, or has a fractal appearance \citep{2009ApJ...696.2086S}. 

The M1 and M2 methods allow the incorporation of new variables for membership analysis, but it is precisely M1 that enables this incorporation in an easier and more elegant way and without preceding hypotheses concerning the variable distribution, except that the cluster members are more densely concentrated than the field stars in that space. 

We have re-determined apparent angular cluster radii directly from the means of the RDPs using the positions data from UCAC4. An analysis of the RDPs has enabled us to correct the central coordinates of 10 open clusters. This work presents, for the first time, a catalogue of stellar clusters whose members have been selected through three different methods. It is a study that we consider fundamental for a subsequent comparison with Gaia data, but which by itself provides unique information, in quantity and variety. From the initial sample, 1876 clusters, the method M1 successfully converges to a solution for a total of 1748 clusters (93\%), M2 for a total of 1693 (90\%) and M3 for a total of 1819 (97\%). The three methods yield membership analysis for a total of 1584 clusters in common (84\%). 

The comparison of the first moments of the proper-motion distribution for the 1584 clusters that have a solution in the three methods corroborates what has been stated. Furthermore, the mean differences among the methods, for the second moments of the cluster member distribution, are lower than the typical proper-motion errors in the UCAC4 catalogue, overall when comparing  M1 and M3. Most of them present differences below the errors catalogued and only a few show clear differences between the estimated values.

We improve the determination of the mean proper motions for the clusters NGC 7092, Ruprecht 147, Mamajek 1, Blanco 1, NGC 752 and Stock 2 for which D14 analysis failed. For these clusters, M1 and M3 obtain satisfactory results which agree with those published by \citet{2001A&A...376..441D}. We have determined  mean proper motions and cluster membership probabilities for 18 open clusters candidates for the first time. 

The results are available in electronic format at SSG-IAA\footnote{http://ssg.iaa.es/en/content/sampedro-cluster-catalog} website as an unique database with two main files. (more information  in Appendix \ref{cat}).

%**********************************************************************
%**********************************************************************
\section*{Acknowledgments}
%**********************************************************************
%**********************************************************************

We thank the referee for his/her comments and suggestions, which have heightened the quality of this work. We warmly thank Sam Lindsay, MNRAS assistant editor, for his invaluable assistance to finish this work in the actual way.  LS and EJA  acknowledge C. Husillos and T. Gallego for helping us with the tailoring of the SSG-IAA database. We acknowledge the IAA-CSIC, Universidade Federal de Itajub\'a (UNIFEI) and IAG/USP for hosting LS during the time this paper was worked on. We acknowledge the financial support from the Spanish Ministry for Economy and Competitiveness and FEDER funds through grants AYA-2010-17631 BES-2011-049077, AYA-2013-40611-P and AYA2016-75931-C2-1-P. LS acknowledges the financial support of the Brazilian funding agency FAPESP (Post-doc fellowship process number 2016/21664-2). AM acknowledges the financial support of the Brazilian funding agency FAPESP (Post-doc fellowship process number 2014/11806-9).  This research has made use of the VizieR catalogue access tool, CDS, Strasbourg, France. The original description of the VizieR service was published in A\&AS 143, 23.

\appendix
%**********************************************************************
%**********************************************************************
\section{}
\label{cat}
%**********************************************************************
%**********************************************************************

Table \ref{catGeneral} describes the parameters in the general catalogue, both for the cluster and the field populations. In addition, it includes  distances, ages and colour excess from DAML02. Parameters related to methods not converging to a solution are set to -9999.99. 

\begin{table*}
        \centering
        \caption{{Description of the parameters included in the general catalogue for the cluster and the field populations. We also provide the distances, ages and color excess from the DAML02 catalogue. The sub-script \textit{i} refers to the three different methods applied in this work.}}
        \vspace{0.2cm}
        \scalebox{1.}[1.]{        
        \label{catGeneral}
        \begin{tabular}{lllll}
        \hline
        \textbf{Position}&  \textbf{Parameter}              &  \textbf{Description}                                                                                &   \textbf{Units} &   \textbf{Type}  \\ 
        \hline
        1                 &        Name                                 & Cluster name                                                                                               &	       --           & String  \\
        2                 &        RA                                      & Right Ascension                                                                                          &	       deg        & Float (5)\\
        3                 &        DEC                                   & Declination                                                                                                   &	       deg        & Float (5)\\
        4                 &        $N_{stars, ini}$                  & Number of stars in the cluster region                                                           &	       --           &Integer\\
        5                 &        $N_{stars}$                        & Number of stars to be analyzed                                                                  &	       --           & Integer\\ 
        6                 &        $N_{Outliers}$                    & Number of Outliers                                                                                     &	       --           & Integer\\
        7, 8, 9         &        $N_{M_i}$                          & Number of cluster members                                                                        &	       --           & Integer\\
        10, 11, 12   &        $P_{M_i}$                           & Percentage of cluster members                                                                  &	 mas/yr         & Float (2)\\
       13, 15, 17    &        $px_{c, M_i}$                     & Cluster mean proper motion ($\mu_{\alpha} cos(\delta)_c$)                       &	 mas/yr         & Float (2)\\
       14, 16, 18    &        $sigpx_{c, M_i}$                 & Cluster proper motion dispersion ($\sigma_{\mu_{\alpha} cos(\delta)_c}$) &	 mas/yr        & Float (2)\\
       19, 21, 23    &        $py_{c, M_i}$                      & Cluster mean proper motion ($\mu_{{\delta}_c}$)                                       &	 mas/yr         & Float (2)\\
       20, 22, 24    &         $sigpy_{c, M_i}$                 & Cluster proper motion dispersion ($\sigma_{{\delta}_c}$)                          &	 mas/yr         & Float (2)\\
       25, 26, 27    &         $coefcorr_{c, M_i}$            & Cluster correlation coefficient  ($\rho_c$)                                                   &	       --           & Float (2)\\
       28                &        Distance                              &  Distance from DAML02                                                                              &	      pc           & Integer\\
       29                &        ColorExcess                        & Color Excess in BV from DAML02                                                              &	       --           & Float (2)\\
       30                &        Age                                      & Age from DAML02 (in log t)                                                                         &	       --           & Float (2)\\
       31                &        Radius                                 & Apparent angular cluster radius from this work                                            &	 arcmin         & Float (2)\\
       32, 34, 36    &        $px_{f, M_i}$                       & Field mean proper motion  ($\mu_{{\alpha} cos(\delta)_f}$)                        &	 mas/yr         & Float (2)\\
       33, 35, 37    &       $sigpx_{f, M_i}$                    & Field proper motion dispersion ($\sigma_{\mu_{\alpha} cos(\delta)_f}$)     &	 mas/yr         & Float (2)\\
       38, 40, 42    &        $py_{f, M_i}$                        & Field mean proper motion ($\mu_{{\delta}_f}$)                                            &	 mas/yr         & Float (2)\\
       39, 41,43    &         $sigpy_{f, M_i}$                   & Field proper motion dispersion ($\sigma_{{\delta}_f}$)                                 &	 mas/yr         & Float (2)\\
       44, 45, 46   &         $coefcorr_{f, M_i}$              & Field correlation coefficient ($\rho_f$)                                                           &	       --           & Float (2)\\
        \hline
        \end{tabular}} 
\end{table*}

We also provide  individual cluster-by-cluster files. These include membership probabilities (for M2 the joint and kinematic probabilities are provided), the outlier classification (outliers are set to 1), three additional columns encoding the membership classification of the methods (cluster members are set to 1 and field stars are set to 0) and additional information from UCAC4.  Membership probabilities for those cases where a method does not converge to a solution are set to -1. An example of an individual cluster-by-cluster file is shown in Table \ref{cat_ngc_2225} for the open cluster NGC 2225.

\begin{table*}
        \centering
        \caption{Example of an individual cluster-by-cluster file for the open cluster NGC 2225. $\alpha$ and $\delta$ are expressed in degrees and $\mu_{\alpha}$ and $\mu_{\delta}$ in mas yr$^{-1}$. $OutFlag$ flags stars classified as outliers (1). $P_{M_i}$ are the membership probabilities obtained by each method. $Class_{M_i}$ flags stars classified as cluster member (1) or field stars (0). Membership probabilities, for those cases where a method does not converge to a solution, are set to -1.}
        \vspace{0.2cm}
        \scalebox{0.93}[0.93]{        
        \label{cat_ngc_2225}
        \begin{tabular}{lclclclclclclclclclclclclclclcl}
        \hline
        \textbf{UCAC4 ID} & $\alpha$ & $\delta$  & $\mu_{\alpha}$  & $\mu_{\delta}$ &  $OutFlag$ & $P_{M_1}$  &  $P_{M_{2,kinem}}$ &  $P_{M_{2,join}}$ &  $P_{M_3}$ & $Class_{M_1}$ &  $Class_{M_2}$ &  $Class_{M_3}$   \\
        \hline
402-013194  & 96.64324  & -9.66397  &  -14.0 &  -3.7  &  0 &  0.01 &  0.86 &  0.87 &  -1. &  0  & 1 &  -1 \\
402-013195  & 96.64333  & -9.62977  &  -19.5 &  29.4 &  0 &  0.00 &  0.28 &  0.67  & -1. &  0  & 0 & -1 \\
402-013199  & 96.64697  & -9.64447  &  56.5  &  -5.8  &  1 &  0.00 &  0.00 &  0.00  & -1. &  0  & 0 &  -1 \\
402-013203  & 96.64961  & -9.67991  &  -58.1 &  -28.5&  1 &  0.00 &  0.00 &  0.00  & -1. &  0  & 0 &  -1 \\
402-013204  & 96.65088  & -9.66370  &  -10.1 &  -28.2&  0 &  0.00 &  0.16 &  0.54  & -1. &  0  & 0 &  -1 \\
402-013211  & 96.65479  & -9.67519  &  -1.4   &  -8.0  &  0 &  0.95  & 0.84 &   0.88 & -1. &  1  & 1 &  -1 \\
402-013214  & 96.65607  & -9.68540  &   -4.2  &  -2.5  &  0 &  0.93  & 0.87 &  0.99  & -1. &  1  & 1 &  -1 \\
402-013216  & 96.65660 & -9.64626   &  -7.5   &  -10.8&  0 &  0.37  & 0.86 &  0.82  & -1. &  0  & 1 &  -1 \\
402-013218  & 96.65835  & -9.61792  &  -29.2 &  3.3   &  0 &  0.00  & 0.25 &  0.00  & -1. &  0  &  0 &  -1 \\
        \hline
        \end{tabular}} 
\end{table*}

\label{lastpage}


\begin{thebibliography}{100}

\bibitem[\protect\citeauthoryear{Alfaro, Cabrera-Cano, \& Delgado}{1991}]{1991ApJ...378..106A} Alfaro E.~J., Cabrera-Cano J., Delgado A.~J., 1991, ApJ, 378, 106
\bibitem[\protect\citeauthoryear{Allison et al.}{2009}]{2009MNRAS.395.1449A} Allison R.~J., Goodwin S.~P., Parker R.~J., Portegies Zwart S.~F., de Grijs R., Kouwenhoven M.~B.~N., 2009, MNRAS, 395, 1449
\bibitem[\protect\citeauthoryear{Bastian, Covey, \& Meyer}{2010}]{2010ARA&A..48..339B} Bastian N., Covey K.~R., Meyer M.~R., 2010, ARA\&A, 48, 339
\bibitem[\protect\citeauthoryear{Baumgardt, Dettbarn, \& Wielen}{2000}]{2000A&AS..146..251B} Baumgardt H., Dettbarn C., Wielen R., 2000, A\&AS, 146, 251
\bibitem[\protect\citeauthoryear{Becker}{1964}]{Becker64} Becker W., 1964, IAUS, 20, 16  
\bibitem[\protect\citeauthoryear{Beshenov \& Loktin}{2004}]{2004A&AT...23..103B} Beshenov G.~V., Loktin A.~V., 2004, A\&AT, 23, 103 
\bibitem[\protect\citeauthoryear{Bonnell \& Davies}{1998}]{1998MNRAS.295..691B} Bonnell I.~A., Davies M.~B., 1998, MNRAS, 295, 691 
\bibitem[\protect\citeauthoryear{Bukowiecki et al.}{2011}]{2011AcA....61..231B} Bukowiecki {\L}., Maciejewski G., Konorski P., Strobel A., 2011, AcA, 61, 231 
\bibitem[\protect\citeauthoryear{Cabrera-Ca\~no \& Alfaro}{1985}]{1985A&A...150..298C} Cabrera-Ca\~no J., Alfaro E.~J., 1985, A\&A, 150, 298 
\bibitem[\protect\citeauthoryear{Cabrera-Ca\~no \& Alfaro}{1990}]{1990A&A...235...94C} Cabrera-Ca\~no, J., Alfaro, E.~J.\ 1990, A\&A, 235, 94
\bibitem[\protect\citeauthoryear{Camargo, Bica, \& Bonatto}{2013}]{2013MNRAS.432.3349C} Camargo D., Bica E., Bonatto C., 2013, MNRAS, 432, 3349 
\bibitem[\protect\citeauthoryear{Charbonnel}{2017}]{2017IAUS..316É.1C} Charbonnel C., 2017, IAUS, 316, 1
\bibitem[\protect\citeauthoryear{Costado et al.}{2016}]{2016arXiv161104398C} Costado M.~T., Alfaro E.~J., Gonzalez M., Sampedro L., 2016, arXiv, arXiv:1611.04398
\bibitem[\protect\citeauthoryear{Dias, L{\'e}pine, \& Alessi}{2001}]{2001A&A...376..441D} Dias W.~S., L{\'e}pine J.~R.~D., Alessi B.~S., 2001, A\&A, 376, 441
\bibitem[\protect\citeauthoryear{Dias et al.}{2002}]{2002A&A...389..871D} Dias W.~S., Alessi B.~S., Moitinho A., L{\'e}pine J.~R.~D., 2002, A\&A, 389, 871
\bibitem[\protect\citeauthoryear{Dias \& L{\'e}pine}{2005}]{2005ApJ...629..825D} Dias W.~S., L{\'e}pine J.~R.~D., 2005, ApJ, 629, 825 
\bibitem[\protect\citeauthoryear{Dias et al.}{2006}]{2006A&A...446..949D} Dias W.~S., Assafin M., Fl{\'o}rio V., Alessi B.~S., L{\'{\i}}bero V., 2006, A\&A, 446, 949 
\bibitem[\protect\citeauthoryear{Dias et al.}{2014}]{2014A&A...564A..79D} Dias W.~S., Monteiro H., Caetano T.~C., L{\'e}pine J.~R.~D., Assafin M., Oliveira A.~F., 2014, A\&A, 564, A79 
\bibitem[\protect\citeauthoryear{Elmegreen et al.}{2000}]{2000prpl.conf..179E} Elmegreen B.~G., Efremov Y., Pudritz R.~E., Zinnecker H., 2000, prpl.conf, 179
\bibitem[\protect\citeauthoryear{Frinchaboy \& Majewski}{2008}]{2008AJ....136..118F} Frinchaboy P.~M., Majewski S.~R., 2008, AJ, 136, 118 
\bibitem[\protect\citeauthoryear{Galadi-Enriquez, Jordi, \& Trullols}{1998}]{1998A&A...337..125G} Galadi-Enriquez D., Jordi C., Trullols E., 1998, A\&A, 337, 125 
\bibitem[\protect\citeauthoryear{Gao}{2016}]{2016RAA....16..184G} Gao X.-H., 2016, RAA, 16, 184 
\bibitem[\protect\citeauthoryear{Gilmore et al. 2012}{}]{2012Msngr.147...25G} Gilmore G., et al., 2012, Msngr, 147, 25 
\bibitem[\protect\citeauthoryear{Heiter et al.}{2014}]{2014A&A...561A..93H} Heiter U., Soubiran C., Netopil M., Paunzen E., 2014, A\&A, 561, A93
\bibitem[\protect\citeauthoryear{H{\o}g et al.}{2000}]{2000A&A...355L..27H} H{\o}g E., et al., 2000, A\&A, 355, L27  
\bibitem[\protect\citeauthoryear{Janes, Tilley, \& Lynga}{1988}]{1988AJ.....95..771J} Janes K.~A., Tilley C., Lynga G., 1988, AJ, 95, 771 
\bibitem[\protect\citeauthoryear{Junqueira et al.}{2015}]{2015MNRAS.449.2336J} Junqueira T.~C., Chiappini C., L{\'e}pine J.~R.~D., Minchev I., Santiago B.~X., 2015, MNRAS, 449, 2336
\bibitem[\protect\citeauthoryear{Kharchenko et al.}{2005}]{2005A&A...438.1163K} Kharchenko N.~V., Piskunov A.~E., R{\"o}ser S., Schilbach E., Scholz R.-D., 2005, A\&A, 438, 1163
\bibitem[\protect\citeauthoryear{Kharchenko et al.}{2012}]{2012A&A...543A.156K} Kharchenko N.~V., Piskunov A.~E., Schilbach E., R{\"o}ser S., Scholz R.-D., 2012, A\&A, 543, A156 
\bibitem[\protect\citeauthoryear{Kharchenko et al.}{2013}]{2013A&A...558A..53K} Kharchenko N.~V., Piskunov A.~E., Schilbach E., R{\"o}ser S., Scholz R.-D., 2013, A\&A, 558, A53
\bibitem[\protect\citeauthoryear{Kroupa}{2001}]{2001MNRAS.322..231K} Kroupa P., 2001, MNRAS, 322, 231
\bibitem[\protect\citeauthoryear{Krumholz et al.}{2014}]{2014prpl.conf..243K} Krumholz M.~R., et al., 2014, prpl.conf, 243 
\bibitem[\protect\citeauthoryear{Lada}{2010}]{2010RSPTA.368..713L} Lada C.~J., 2010, RSPTA, 368, 713 
\bibitem[\protect\citeauthoryear{Landin et al.}{2006}]{2006A&A...456..269L} Landin N.~R., Ventura P., D'Antona F., Mendes L.~T.~S., Vaz L.~P.~R., 2006, A\&A, 456, 269 
\bibitem[\protect\citeauthoryear{Larson}{1994}]{1994LNP...439...13L} Larson R.~B., 1994, LNP, 439, 13
\bibitem[\protect\citeauthoryear{L{\'e}pine et al.}{2011}]{2011MNRAS.417..698L} L{\'e}pine J.~R.~D., et al., 2011, MNRAS, 417, 698 
\bibitem[\protect\citeauthoryear{Loktin \& Beshenov}{2003}]{2003ARep...47....6L} Loktin A.~V., Beshenov G.~V., 2003, ARep, 47, 6 
\bibitem[\protect\citeauthoryear{Magrini et al.}{2009}]{2009A&A...494...95M} Magrini L., Sestito P., Randich S., Galli D., 2009, A\&A, 494, 95
\bibitem[\protect\citeauthoryear{Magrini et al.}{2017}]{2017arXiv170300762M} Magrini L., et al., 2017, arXiv, arXiv:1703.00762 
\bibitem[\protect\citeauthoryear{Ochsenbein, Bauer, \& Marcout}{2000}]{2000A&AS..143...23O} Ochsenbein F., Bauer P., Marcout J., 2000, A\&AS, 143, 23 
\bibitem[\protect\citeauthoryear{Oliveira et al.}{2013}]{2013A&A...557A..14O} Oliveira A.~F., Monteiro H., Dias W.~S., Caetano T.~C., 2013, A\&A, 557, A14 
\bibitem[\protect\citeauthoryear{Parker \& Dale}{2017}]{Parker17} Parker R.~J., Dale J.~E., 2017, arXiv, arXiv:1705.04686 
\bibitem[\protect\citeauthoryear{Parker, Goodwin, \& Allison}{2011}]{Parker11} Parker R.~J., Goodwin S.~P., Allison R.~J., 2011, MNRAS, 418, 2565 
\bibitem[\protect\citeauthoryear{Parker et al.}{2009}]{Parker09} Parker R.~J., Goodwin S.~P., Kroupa P., Kouwenhoven M.~B.~N., 2009, MNRAS, 397, 1577
\bibitem[\protect\citeauthoryear{Piatti, Dias, \& Sampedro}{2017}]{2017MNRAS.466..392P} Piatti A.~E., Dias W.~S., Sampedro L.~M., 2017, MNRAS, 466, 392
\bibitem[\protect\citeauthoryear{Roeser, Demleitner, \& Schilbach}{2010}]{2010AJ....139.2440R} Roeser S., Demleitner M., Schilbach E., 2010, AJ, 139, 2440
\bibitem[\protect\citeauthoryear{Sampedro \& Alfaro}{2016}]{2016MNRAS.457.3949S} Sampedro L., Alfaro E.~J., 2016, MNRAS, 457, 3949 
\bibitem[\protect\citeauthoryear{S{\'a}nchez \& Alfaro}{2009}]{2009ApJ...696.2086S} S{\'a}nchez N., Alfaro E.~J., 2009, ApJ, 696, 2086 
\bibitem[\protect\citeauthoryear{S{\'a}nchez, Vicente, \& Alfaro}{2010}]{2010A&A...510A..78S} S{\'a}nchez N., Vicente B., Alfaro E.~J., 2010, A\&A, 510, A78 

\bibitem[\protect\citeauthoryear{Kulkarni \& Harman}{2011}]{bayes_rule} Sanjeev Kulkarni \& Gilbert Harman, 2011, Elementary Introduction to Statistical Learning Theory. John Wiley \& Sons, Inc., Hoboken, New Jersey 
\bibitem[\protect\citeauthoryear{Sharma et al.}{2006}]{2006AJ....132.1669S} Sharma S., Pandey A.~K., Ogura K., Mito H., Tarusawa K., Sagar R., 2006, AJ, 132, 1669 
\bibitem[\protect\citeauthoryear{Skrutskie et al.}{2006}]{2006AJ....131.1163S} Skrutskie M.~F., et al., 2006, AJ, 131, 1163 
\bibitem[\protect\citeauthoryear{van den Bergh \& McClure}{1980}]{1980A&A....88..360V} van den Bergh S., McClure R.~D., 1980, A\&A, 88, 360
\bibitem[\protect\citeauthoryear{Vasilevskis, Klemola, \& Preston}{1958}]{1958AJ.....63..387V} Vasilevskis S., Klemola A., Preston G., 1958, AJ, 63, 387 

\bibitem[\protect\citeauthoryear{Vicente, S{\'a}nchez, \& Alfaro}{2016}]{2016MNRAS.461.2519V} Vicente B., S{\'a}nchez N., Alfaro E.~J., 2016, MNRAS, 461, 2519
\bibitem[\protect\citeauthoryear{Wolfe}{1970}]{Wolfe}Wolfe, J. H. 1970, Multivariate Behavioral Research, 5, 329, 350.

\bibitem[\protect\citeauthoryear{W{\"u}nsch et al.}{2017}]{2017ApJ...835...60W} W{\"u}nsch R., Palou{\v s} J., Tenorio-Tagle G., Ehlerov{\'a} S., 2017, ApJ, 835, 60 
\bibitem[\protect\citeauthoryear{Zacharias et al.}{2013}]{2013AJ....145...44Z} Zacharias N., Finch C.~T., Girard T.~M., Henden A., Bartlett J.~L., Monet D.~G., Zacharias M.~I., 2013, AJ, 145, 44 
\bibitem[\protect\citeauthoryear{Zhao \& He}{1990}]{1990A&A...237...54Z} Zhao J.~L., He Y.~P., 1990, A\&A, 237, 54

\end{thebibliography}
\end{document}